\documentstyle[12pt]{article}
\newcommand{\cd}{\makebox[0.08cm]{$\cdot$}}
\title
{\bf {The nucleon wave function in light-front dynamics}}
 \author{ V.A. Karmanov\thanks{e-mail: karmanov@sci.lebedev.ru}
\\{\small \em  Lebedev Physical Institute, Leninsky Prospekt 53, 117924
Moscow, Russia}}

\begin{document}
\maketitle
\begin{center}
{\large Submitted to Nucl. Phys. A}
\end{center}
\vspace{0.5cm}

\begin{abstract}
The general spin structure of the relativistic nucleon wave function in 
the $3q$-model is found. It contains 16 spin components, in contrast to 
8 ones known previously, since in a many-body system the parity 
conservation does not reduce the number of the components.  The 
explicitly covariant form of the wave function automatically takes into 
account the relativistic spin rotations, without introducing any Melosh 
rotation matrices. It also reduces the calculations to the standard 
routine of the Dirac matrices and of the trace techniques.  In examples 
of the proton magnetic moment and of the axial nucleon form factor, 
with a particular wave function, we reproduce the results of the 
standard approach. Calculations beyond the standard assumptions give 
different results.  
\end{abstract}

\section{Introduction}\label{sec1}
The light-front dynamics (LFD) is a powerful approach to the theory of 
relativistic composite systems. In many papers, (see, for example, 
\cite{bt76}-\cite{salme95}) it was applied to the relativistic 
three-body systems, and, in particular, to the nucleon in $3q$-model 
(see for review \cite{kp91,coester92}).  In the paper \cite{dz88} eight 
spin components of the nucleon wave function were indicated. Majority 
of calculations was done with one of the components only, corresponding 
to the fully symmetric momentum independent (S-wave) spin structure.

The aim of the present paper is two-fold.  First of all, we discover 
that eight components do not exhaust the relativistic nucleon wave 
function.  We will find another eight components, so the total number 
of them is sixteen.  In general, this is related to the fact known long 
ago \cite{kolyb65} that in a many-body system the parity conservation 
does not reduce the number of the spin components. Hence, for the 
nucleon we get $2\times 2\times 2\times 2=16$. This opportunity is 
absent in any two-body system and in the nonrelativistic three-body 
one. Technically, the extra components can be constructed since the 
particle four-momenta in any off-energy-shell relativistic amplitude, 
in particular, in the wave function, are not related by the 
conservation law: for the minus-projections $k_-=k_0-k_z$ the sum of 
the quark momenta $k_{1,2,3}$ and the nucleon momentum $p$ are not 
equal to each other:  $(k_1+k_2+k_3)_-\neq p_-$. Hence, we have in our 
disposal 4 four-vectors and can construct the pseudoscalar 
$C_{ps}=e^{\mu\nu\rho\gamma}k_{1\mu}k_{2\nu}k_{3\rho}p_{\gamma}$.  Due 
to that, in addition to ``old" eight components given in \cite{dz88}, 
we construct another eight spin structures with ``wrong" P-parity (the 
pseudoscalar structures) and then ``correct" them by multiplying by 
$C_{ps}$. However, this way to find the spin components is not 
obligatory.  One can construct an equivalent set of sixteen components 
such that only a few of them (less than eight) contain the factor 
$C_{ps}$. These extra components (relative to the paper \cite{dz88}) 
are necessary in order to represent even symmetric S-wave spin 
structure (initially given in c.m.-system) in arbitrary system of 
reference in terms of the Dirac matrices sandwiched with the spinors.  
In other language this corresponds to multiplying the center-of-mass 
wave function by the Melosh rotation matrices.  If we will omit these 
extra components and come back to the c.m.-system, we would not 
reproduce our initial S-wave but again will find some extra components. 
In general case one should start with the wave function containing all 
sixteen components in any system of reference. They are forming the 
full basis.  Their total number does not depends, of course, on the 
representation.  Their relative magnitude is determined by dynamics.

Secondly, we will represent the nucleon wave function in the $3q$ model 
in the explicitly covariant form. This will allow one to use in 
calculations the standard Dirac-matrices algebra and the trace 
techniques.  In particular, we will see that there is no any need to 
introduce explicitly any Melosh rotation matrices: the covariant 
approach incorporates automatically the spin rotation effects.  In 
examples of the proton magnetic moment and of the axial nucleon form 
factor, we reproduce by this way the results which in the standard 
approach are obtained due to averaging the Melosh matrices. 

The above problems will be solved in the explicitly covariant version 
of LFD developed in a series of papers, starting with \cite{karm76} 
(see for review \cite{cdkm}). In the standard LFD the wave functions 
are defined on the light-front plane $t+z=0$. In the covariant approach 
the wave functions are defined on the light-front plane given by the 
invariant equation $\omega\cd x=0$, where $\omega$ is the four-vector 
with $\omega^2=0$. This provides all the advantages of the explicit 
covariance, similar to advantages of the Feynman graph techniques over 
the old fashioned perturbation theory.  The standard approach is 
obtained as a particular case at $\omega= (1,0,0,-1)$.  However, we 
would like to emphasize that this particular choice of $\omega$ does 
not reduce the number of the components of the nucleon wave function, 
it is 16 in any version of LFD.

Plan of the paper is the following.  In sect. \ref{covwf} we introduce 
special representation for the wave function and the three-dimensional 
variables with the simple transformation properties. In sect. \ref{so} 
we construct the corresponding spin matrices with the same 
transformation properties.  In sect. \ref{spst} sixteen spin structures 
of the nucleon wave function are found, however without taking into 
account any permutation symmetry.  In sect.  \ref{pg} we remind the 
permutation group properties in the three-body case. In sect. 
\ref{indep} we take into account the appropriate permutation symmetries 
of the spin structures independent of momenta.  Sect.  \ref{depen} is 
devoted to incorporating the permutation symmetry in the components 
depending on momenta. In sect. \ref{meq} the matrix elements of the 
current operator, used in the form factor calculations, are found. In 
sect. \ref{appl} we consider the examples of the proton magnetic moment 
and of the axial nucleon form factor. Section \ref{concl} contains the 
concluding remarks.

\section{The covariant light-front wave function}\label{covwf}
The properties of the covariant light-front wave function are described 
in \cite{cdkm}. Here we remind some of them in the three-body case.

The light front wave function of the nucleon composed from three quarks 
has the form:
\begin{equation}\label{eqn1}
{\mit\Phi}={\mit\Phi}^{\sigma}_{\sigma_1\sigma_2\sigma_3}
(k_1,k_2,k_3,p, \omega\tau),
\end{equation}
where $k_{1-3}$ and $p$ are the quark and the nucleon four-momenta, 
$\tau$ is a scalar parameter and $\sigma_{1-3}$, $\sigma$ are the quark 
and nucleon spin projections on $z$-axis in the corresponding frames of 
the rest.  The four-momenta are related by the conservation law:  
\begin{equation}\label{eq29a} 
p+\omega\tau=k_1+k_2+k_3 \equiv {\cal P}.  
\end{equation} 
The presence in (\ref{eq29a}) of the term $\omega\tau$ just reflects 
the nonconservation of the minus-components, since for 
$\omega=(1,0,0,-1)$ this term contributes only to the minus-projection 
of this equation.

Below we will use the effective mass corresponding to the four-momentum 
${\cal P}$:
\begin{equation}\label{mcal} 
{\cal M}=\sqrt{{\cal P}^2}.
\end{equation} 

Under rotations and the Lorentz transformations $g$ the wave 
function (\ref{eqn1}) is transformed by the rotation matrices 
$D^{(\frac{1}{2})}\{R\}$ depending on the different rotation operators 
$R$ for the particles with different momenta.  Namely, 
\begin{eqnarray}\label{wfp6} 
&&{\mit\Phi}^{ \sigma}_{\sigma_1 \sigma_2\sigma_3} 
(gk_1,gk_2,gk_3,gp,g\omega\tau) 
=\sum_{\sigma'\sigma'_1\sigma'_2\sigma'_3} 
D^{(\frac{1}{2})*}_{\sigma\sigma'}\{R[g,p]\} 
\nonumber \\ 
&&\times
D^{(\frac{1}{2})}_{\sigma_1\sigma'_1}\{R[g,k_1]\} 
D^{(\frac{1}{2})}_{\sigma_2\sigma'_2}\{R[g,k_2]\} 
D^{(\frac{1}{2})}_{\sigma_3\sigma'_3}\{R[g,k_3]\} 
{\mit\Phi}^{\sigma'}_{\sigma'_1 
\sigma'_2\sigma'_3}(k_1,k_2,k_3,p,\omega\tau),
\end{eqnarray} 
where, for example, $R[g,p]$ is the following rotation operator:
\begin{equation}\label{rot}                                                     
R[g,p]=L^{-1}(gp)gL(p).
\end{equation}                                                                  
$L(p)$ is the Lorentz boost corresponding to the velocity 
$\vec{v}=\vec{p}/p_0$.  Below, in addition to the wave function ${\mit 
\Phi}$ transformed by (\ref{wfp6}), we will introduce another 
representation in which the wave function, denoted as ${\mit \Psi}$, is 
transformed by one and the same rotation for all the particles. This 
representation is defined as (see \cite{karm79}):  
\begin{eqnarray}\label{sp1} 
&&{\mit\Psi}^{\sigma}_{\sigma_1 \sigma_2 
\sigma_3}(k_1,k_2,k_3,p,\omega\tau) = 
\sum_{\sigma',\sigma_1',\sigma_2',\sigma_3'} 
D^{(\frac{1}{2})*}_{\sigma\sigma'}\{R[L^{-1}({\cal P}), p]\} 
D^{(\frac{1}{2})}_{\sigma_1\sigma_1'}\{R[L^{-1}({\cal P}),k_1]\} 
\nonumber \\ &&\times 
D^{(\frac{1}{2})}_{\sigma_2\sigma_2'}\{R[L^{-1}({\cal P}),k_2]\} 
D^{(\frac{1}{2})}_{\sigma_3\sigma_3'}\{R[L^{-1}({\cal P}),k_3]\} 
{\mit\Phi}^{\sigma'}_{\sigma_1' \sigma_2' \sigma_2'} 
(k_1,k_2,k_3,p,\omega\tau)\ , 
\end{eqnarray} 
where, e.g.,  
$R[L^{-1}({\cal P}),p]$ is given by (\ref{rot}) with $g=L^{-1}({\cal 
P})$.  The matrix $D^{(\frac{1}{2})}_{\sigma_1\sigma'_1} 
\{R[L^{-1}({\cal P}),k_1]\}$, appeared in the transformation 
(\ref{sp1}), has the form \cite{shir54}:  
\begin{equation}\label{nz4} 
D^{\frac{1}{2}}\{R[L^{-1}({\cal P}),k_1]\} = \frac{(k_{10}+m)({\cal 
P}_0 + {\cal M}) - \vec{\sigma}\cd\vec{{\cal P}}\;
\vec{\sigma}\cd\vec{k_1}} {[2(k_{10}+m)({\cal P}_0 + {\cal M})
(k_1\cd{\cal P} + m{\cal M})]^{1/2}},
\end{equation} 
and similarly for other matrices. We will not use their explicit form 
below.

In this representation the wave function is transformed as 
\cite{karm79}:
\begin{eqnarray}\label{sp4}                                                     
&&{\mit\Psi}^{\sigma}_{\sigma_1\sigma_2\sigma_3}
(gk_1,gk_2,gk_3,gp,g\omega\tau) 
 =  \sum_{\sigma',\sigma'_1\sigma'_2\sigma'_3} 
D^{(\frac{1}{2})*}_{\sigma\sigma'}\{R[g,{\cal P}]\} 
\nonumber \\ 
&&\times
D^{(\frac{1}{2})}_{\sigma_1\sigma'_1}\{R[g,{\cal P}]\} 
D^{(\frac{1}{2})}_{\sigma_2\sigma'_2}\{R[g,{\cal P}]\} 
D^{(\frac{1}{2})}_{\sigma_3\sigma'_3}\{R[g,{\cal P}]\} 
{\mit\Psi}^{\sigma'}_{\sigma'_1\sigma'_2\sigma'_3}
(k_1,k_2,k_3,p,\omega\tau)\ .  
\end{eqnarray}                                                                  
All the matrices $D^{(\frac{1}{2})}\{R[g,{\cal P}]\}$ in (\ref{sp4}) 
depend on the same rotation operator $R[g,{\cal P}]$.  

Instead of the four-vectors $k_{1-3},\omega$ we introduce the 
three-vector variables which are constructed as follows:
\begin{equation}\label{sp5}
\vec{q}_i=L^{-1}({\cal P})\vec{k}_i
= \vec{k}_i -                                
\frac{\vec{\cal P}}{\cal M}\left[k_{i0} - 
\frac{\vec{k}_i\cd\vec{{\cal P}}}{{\cal M}+{\cal P}_0}\right]\ , 
\end{equation}                                                                  
\begin{equation}\label{sc5}                                                     
\vec{n} = L^{-1}({\cal P})\vec{\omega}/|L^{-1}({\cal P})                        
\vec{\omega}|, 
\end{equation}
$i=1,2,3$.
Note that $\vec{q}_1+ \vec{q}_2+ \vec{q}_3 =0.$  These variables are 
transformed by the same rotation operator $R[g,{\cal P}]$ which 
transforms the spin projections in (\ref{sp4}):
\begin{equation}\label{sp6}
\vec{q}_i\,'=R[g,{\cal P}]\vec{q}_i,
\quad \vec{n}\,'=R[g,{\cal P}]\vec{n}.
\end{equation}
The same situation takes place in the nonrelativistic case: all the 
spins and momenta are transformed by one and the same rotation.  
Therefore in the representation (\ref{sp1}) and in the variables 
(\ref{sp5}), (\ref{sc5}) the problem of constructing the general form 
of the nucleon wave function is analogous to the nonrelativistic one. 
This fact simplifies very much the construction of the spin states. The 
only difference is the presence of the extra vector $\vec{n}$, due to 
dependence of the wave function on the orientation of the light-front 
plane.  We will need also the spin operators which are transformed 
similarly to $\vec{q}_i,\vec{n}$, eq. (\ref{sp6}). These operators will 
be constructed in the next section.

The transformation (\ref{sp1}) is similar to the Melosh transformation 
\cite{melosh}.  However, the wave function 
${\mit\Phi}^{\sigma}_{\sigma_1\sigma_2\sigma_3}$, eq. (\ref{eqn1}), 
does not require any Melosh matrices. It can be constructed in terms of 
the Dirac spinors, that automatically provides the correct coupling of 
the quark spin and angular momenta in the nucleon spin. We will come 
back to this point below.

In the system of reference where $\vec{\cal P}=0$ the wave functions in 
two representations coincide with each other. Therefore, to establish 
the relation between the wave functions in two representations we do 
not need the explicit form of the rotation matrices in the above 
formulas.  It is enough to compare the wave functions in the system 
where $\vec{\cal P}=0$.

\section{The three-dimensional spin matrices}\label{so}
In this section we construct the three-dimensional spin matrices 
which ({\it i}) are transformed similarly to the vectors 
$\vec{q}_i,\vec{n}$, eq. (\ref{sp6}), and ({\it ii}) in the system 
where $\vec{\cal P}=0$ turn into the Pauli matrices. Then in terms of 
these operators and of the vectors $\vec{q}_i,\vec{n}$ we will 
construct all the spin structures of the nucleon wave function. With 
these matrices the problem of constructing the representations of the 
permutation group is also simplified very much.

The point ({\it i}) is fulfilled in the representation (\ref{sp1}), 
the point ({\it ii}) is provided by the projection operators 
constructed below.

We introduce the spinor $\bar{u}^{\sigma_1}_{\cal P}(k_1)$ in the 
representation (\ref{sp1}):  
\begin{equation}\label{sp7}
\bar{u}^{\sigma_1}_{\cal P}(k_1)=\sum_{\sigma'_1}
D^{(\frac{1}{2})}_{\sigma_1\sigma_1'}\{R[L^{-1}({\cal P}),k_1]\} 
\bar{u}^{\sigma'_1}(k_1),
\end{equation}
and similarly for other spinors. The explicit form of the spinor
$\bar{u}^{\sigma_1}(k_1) $ and of the other ones is given in appendix 
\ref{appen1}.

We construct also the projection operators:  
\begin{equation}\label{eq77} 
\Pi_+=\frac{{\cal M}+\hat{\cal P}}{2{\cal M}},\quad
\Pi_-=\frac{{\cal M}-\hat{\cal P}}{2{\cal M}},
\end{equation}
${\cal P}$ is defined by (\ref{eq29a}),
${\cal M}$ is defined by (\ref{mcal}) and 
$\hat{\cal P}=\gamma^{\mu}{\cal P}_{\mu}$.
They have the properties: 
$$
\Pi_{\pm}^2=\Pi_{\pm},\quad \Pi_+\Pi_-=\Pi_-\Pi_+=0, \quad
\gamma_5\Pi_+=\Pi_-\gamma_5,\quad
U_c\Pi_+^t=\Pi_- U_c
$$
(the index $t$ means the transposition).  In the system of reference 
where $\vec{\cal P}=0$, ${\cal P}_0={\cal M}$, the projection operators 
obtain the simple form:
\begin{equation}\label{eq80}
\left.\Pi_+\right|_{\vec{\cal P}=0}=\frac{1}{2} (1+\gamma^0)=
\left(
\begin{array}{cc}
1&0\\ 0&0
\end{array}\right),\quad
\left.\Pi_-\right|_{\vec{\cal P}=0}=\frac{1}{2} (1-\gamma^0)=
\left(
\begin{array}{cc}
0&0\\ 0&1 
\end{array}\right).
\end{equation}
Then the spin matrices we are looking for are constructed as follows:
\begin{eqnarray}\label{eq80n}
\vec{\sigma}_{12}&=&-c_1c_2\bar{u}_{\cal P}^{\sigma_1}(k_1)
\Pi_+\vec{\gamma}\Pi_-U_c\bar{u}^{\sigma_2}_{\cal P}(k_2),
\nonumber\\
\vec{\sigma}_{3N}&=&-c_3c_N\bar{u}_{\cal P}^{\sigma_3}(k_3)
\Pi_+\vec{\gamma}\gamma_5\Pi_+u^{\sigma}_{\cal P}(p),
\end{eqnarray}
where $U_c=\gamma^2\gamma^0$ is the charge conjugation matrix,
\begin{equation}\label{eq81a} 
c_{1,2,3}= \frac{1}{\sqrt{\varepsilon_{q_{1,2,3}}+m}},\quad c_N= 
\frac{1}{\sqrt{\varepsilon_{p}+M}}.  
\end{equation} 
The energies in (\ref{eq81a}) are defined in the system where 
 $\vec{\cal P}=0$. They are expressed through the invariants:  $$ 
\varepsilon_{q_i}=\sqrt{\vec{q}_i\,^2+m^2}={\cal P}\cd k_i/{\cal M},
\quad 
\varepsilon_p={\cal P}\cd p/{\cal M}.
$$ 
In eqs. (\ref{eq80n}) the quantities $\vec{\sigma}_{12}$, 
$\vec{\sigma}_{3N}$ are the $2\times 2$ - matrices relative to the 
indices $\sigma_1,\sigma_2$ and $\sigma_3,\sigma$.  

The matrices $\vec{\sigma}_{12}, \vec{\sigma}_{3N}$ are transformed by 
(\ref{sp4}), like all the amplitudes and the variables in the 
representation (\ref{sp1}).  In the system of reference where 
$\vec{\cal P}=0$, with the explicit form of the spinors given in 
appendix \ref{appen1} (eqs.(\ref{eq74})-(\ref{eq76})) and with the 
projections operators (\ref{eq80}), we find:  
\begin{eqnarray}\label{sp10} 
\left.c_1\bar{u}(k_1)\Pi_+\right|_{\vec{\cal P}=0}=w^\dagger_1(1,0), 
\quad  
\left.c_3\bar{u}(k_3)\Pi_+\right|_{\vec{\cal P}=0}=w^\dagger_3(1,0),
\nonumber\\
\left.c_2\Pi_-U_c\bar{u}^t(k_2)\right|_{\vec{\cal P}=0}=\left(
\begin{array}{r}
0\\
-1
\end{array} \right)\sigma_yw^*_2,
\quad
\left.c_N\Pi_+u(p)\right|_{\vec{\cal P}=0}=\left(
\begin{array}{r}
1\\
0
\end{array} \right)w_N,
\end{eqnarray}
$w$'s are the two-components spinors. The factors $c_i$, eqs. 
(\ref{eq81a}), just cancel the factors $\sqrt{\varepsilon +m}$ in the 
bi-spinors.

With the explicit form of the Dirac matrices, eqs. (\ref{dg}), we see 
that at $\vec{\cal P}=0$ the operators 
$\vec{\sigma}_{12},\vec{\sigma}_{3N}$ indeed coincide with the Pauli 
matrices:
\begin{eqnarray}\label{eq80nn}
\vec{\sigma}_{12}|_{\vec{\cal P}=0}&=&
(w_1^{\dagger}\vec{\sigma}\sigma_y w^*_2),
\nonumber\\
\vec{\sigma}_{3N}|_{\vec{\cal P}=0}&=&
(w_3^{\dagger}\vec{\sigma} w_N).
\end{eqnarray}
They correspond to coupling of  spins of the  pairs in the spin 1.

We also introduce the unit operators 
\begin{eqnarray}\label{eq35a} 
1_{12} &=& \left.c_1c_2\bar{u}_{\cal P}^{\sigma_1}(k_1)
\Pi_+\gamma_5\Pi_-U_c\bar{u}^{\sigma_2}_{\cal P}(k_2) 
\right|_{\vec{\cal P}=0}=
(w_1^{\dagger}\sigma_y w^*_2),
\nonumber\\
1_{3N} &=& \left.c_3c_N\bar{u}_{\cal P}^{\sigma_3}(k_3)
\Pi_+\Pi_+u^{\sigma}_{\cal P}(p)\right|_{\vec{\cal P}=0}=
(w_3^{\dagger} w_N)
\end{eqnarray}
corresponding to the zero pair spins.  Below for shortness we will 
often replace these operators by 1 without indices.

\section{The spin  structure of the nucleon wave function}\label{spst}
As mentioned, in the variables $\vec{q}_1,\vec{q}_2,\vec{q}_3,\vec{n}$ 
the problem of finding the spin structures of the nucleon wave function 
coincides with the nonrelativistic one, since these variables are 
subjected to the rotations only. The only difference is the presence of 
the extra vector $\vec{n}$. Formally it coincides with the problem of 
decomposition in the invariant amplitudes of the reaction amplitude 
$1+2\rightarrow 3+4+5$, since the wave function (\ref{eqn1}), like the 
amplitude of a process $1+2\rightarrow 3+4+5$, depends on five 
four-vectors satisfying the conservation law (\ref{eq29a}).
In general case, for the nonrelativistic reaction $1+2\rightarrow 
3+\ldots+n$ this problem was solved in the paper \cite{kolyb65}.  It 
was shown that starting with $n=5$ the number of structures is the 
product of the factors $(2j+1)$, corresponding to spins of all the 
particles. This gives 16 for nucleon. Below we will find these 16 
structures of the nucleon wave function and then take into account the 
requirements of the permutation group.

First of all, we have two structures independent of momenta,
the unit operator and the scalar product of $\vec{\sigma}$'s:
$$
1\equiv 1_{12} 1_{3N}, \quad
\mbox{and} \quad \vec{\sigma}_{12}\cd\vec{\sigma}_{3N}.
$$
Then we construct the tensor of the second rank 
$$
\sigma_{12}^i \sigma_{3N}^j.
$$
It will be contracted with the corresponding tensors in the momentum 
space. The latters are constructed as follows.

We start with the set of ten tensors symmetric in the indices $i,j$:
\begin{eqnarray}\label{eq46}
&&T^{ij}_{11}=q_{1i}q_{1j}-\frac{1}{3}\vec{q}\,^2_1\delta_{ij},\quad
T^{ij}_{22}=q_{2i}q_{2j}-\frac{1}{3}\vec{q}\,^2_2\delta_{ij},\quad
T^{ij}_{33}=q_{3i}q_{3j}-\frac{1}{3}\vec{q}\,^2_3\delta_{ij};
\nonumber\\
&&
\begin{array}{ll}
T^{ij}_{12}=q_{1i}q_{2j}+q_{2i}q_{1j}
-\frac{\displaystyle{2}}{\displaystyle{3}}
\vec{q}_1\cd\vec{q}_2\delta_{ij},
&T^{ij}_{13}=q_{1i}q_{3j}+q_{3i}q_{1j}
-\frac{\displaystyle{2}}{\displaystyle{3}}
\vec{q}_1\cd\vec{q}_3\delta_{ij},
\\
&T^{ij}_{23}=q_{2i}q_{3j}+q_{3i}q_{2j}
-\frac{\displaystyle{2}}{\displaystyle{3}}
\vec{q}_2\cd\vec{q}_3\delta_{ij};
\\
T^{ij}_{1n}=q_{1i}n_{j}+n_{i}q_{1j}
-\frac{\displaystyle{2}}{\displaystyle{3}}
\vec{q}_1\cd\vec{n}\delta_{ij}, 
&T^{ij}_{2n}=q_{2i}n_{j}+n_{i}q_{2j}
-\frac{\displaystyle{2}}{\displaystyle{3}}
\vec{q}_2\cd\vec{n}\delta_{ij}, 
\\
&T^{ij}_{3n}=q_{3i}n_{j}+n_{i}q_{3j}
-\frac{\displaystyle{2}}{\displaystyle{3}}
\vec{q}_3\cd\vec{n}\delta_{ij};
\\
T^{ij}_{nn}=n_{i}n_{j}-\frac{\displaystyle{1}}
{\displaystyle{3}}\delta_{ij}.&
\end{array}
\end{eqnarray}
Since $\vec{q}_1+\vec{q}_2+\vec{q}_3=0$, the tensors in eq. 
(\ref{eq46}) are not independent from each other. We will express the 
nondiagonal tensors $T^{ij}_{12},T^{ij}_{13},T^{ij}_{23}$ through the 
diagonal ones.  Namely:  
\begin{eqnarray}\label{eq46a} 
&&T_{12}=T_{33}-T_{11}-T_{22}
\nonumber\\
&&T_{13}=T_{22}-T_{11}-T_{33}
\nonumber\\
&&T_{23}=T_{11}-T_{22}-T_{33}
\end{eqnarray}
The tensors $T_{1n},T_{2n},T_{3n}$ are also related with each other:  
\begin{equation}\label{eq46b}
T_{1n} + T_{2n} + T_{3n}=0. 
\end{equation}  
Four conditions (\ref{eq46a}), (\ref{eq46b}) reduce the number of 
tensors down to six. However, since the symmetric traceless tensor have 
five components, there should be only five independent tensors and, 
hence, a relation between these six tensors. This relation is given in 
appendix \ref{appen2} by eq. (\ref{eq31}). Due to this relation we 
exclude the tensor $T^{ij}_{nn}$ which does not contain the relative 
momenta. Hence, we have five independent structures to be contracted 
with the spin tensor $\sigma_{12}^i \sigma_{3N}^j$.

Then we construct four tensors, antisymmetric in the indices $i,j$:
\begin{eqnarray}\label{eq60}
A_{12}&=&q_{1i}q_{2j}-q_{1j}q_{2i}
\nonumber\\
A_{1n}&=&q_{1i}n_j -q_{1j}n_i
\nonumber\\
A_{2n}&=&q_{2i}n_j -q_{2j}n_i
\nonumber\\
A_{3n}&=&q_{3i}n_j -q_{3j}n_i
\end{eqnarray}
Since $A_{1n}+A_{2n}+A_{3n}=0$, there are three independent tensors.  
We will specify them later.

We can also construct the linear structures of the type 
$\vec{\sigma}\cd [\vec{q}_1\times\vec{q}_2]$ (multiplied implicitly 
by the corresponding unit operator).  By this way we find sixteen spin 
structures of the nucleon wave function:
\begin{eqnarray}
({\it 1-2}) && 1,\quad
 \vec{\sigma}_{12}\cd  \vec{\sigma}_{3N};
\nonumber\\
({\it 3-5}) && T_{11}^{ij}\sigma_{12}^i\sigma_{3N}^j,\quad
T_{22}^{ij}\sigma_{12}^i\sigma_{3N}^j,\quad
T_{33}^{ij}\sigma_{12}^i\sigma_{3N}^j;
\nonumber\\
({\it 6-7}) && \mbox{two structures of three:}
\nonumber\\
&&T_{1n}^{ij}\sigma_{12}^i\sigma_{3N}^j,\quad
T_{2n}^{ij}\sigma_{12}^i\sigma_{3N}^j,\quad
T_{3n}^{ij}\sigma_{12}^i\sigma_{3N}^j;
\nonumber\\
({\it 8})&& A_{12}^{ij}\sigma_{12}^i\sigma_{3N}^j;
\nonumber\\
({\it 9-10}) &&\mbox{two structures of three:}
\nonumber\\
&&A_{1n}^{ij}\sigma_{12}^i\sigma_{3N}^j,\quad
A_{2n}^{ij}\sigma_{12}^i\sigma_{3N}^j,\quad
A_{3n}^{ij}\sigma_{12}^i\sigma_{3N}^j; 
\nonumber\\
({\it 11-16}) &&\vec{\sigma}_{12}\cd[\vec{q}_1\times\vec{q}_2],\quad
\vec{\sigma}_{12}\cd[\vec{q}_1\times\vec{n}],\quad
\vec{\sigma}_{12}\cd[\vec{q}_2\times\vec{n}],
\nonumber\\
&&\vec{\sigma}_{3N}\cd[\vec{q}_1\times\vec{q}_2],\quad
\vec{\sigma}_{3N}\cd[\vec{q}_1\times\vec{n}],\quad
\vec{\sigma}_{3N}\cd[\vec{q}_2\times\vec{n}].
\label{eq34c}
\end{eqnarray}
Instead of the structures ({\it 8-10}) one can construct 
$$([\vec{\sigma}_{12}\times \vec{\sigma}_{3N}]\cd \vec{q}_1) 
(\vec{n}\cd[\vec{q}_1\times\vec{q}_2]),\quad
([\vec{\sigma}_{12}\times \vec{\sigma}_{3N}]\cd \vec{q}_2) 
(\vec{n}\cd[\vec{q}_1\times\vec{q}_2]),\quad
([\vec{\sigma}_{12}\times \vec{\sigma}_{3N}]\cd \vec{n}) 
(\vec{n}\cd[\vec{q}_2\times\vec{q}_2]),$$ 
but the double vector 
products are reduced to the linear combinations of ({\it 8-10}). 
Instead of the structures ({\it 11-16}) one can construct 
$(\vec{\sigma}_{12}\cd\vec{q}_1) 
(\vec{n}\cd[\vec{q}_1\times\vec{q}_2])$, 
$(\vec{\sigma}_{12}\cd\vec{q}_2) 
(\vec{n}\cd[\vec{q}_1\times\vec{q}_2])$, etc., but they are linearly 
expressed through the structures ({\it 11-16}) by the formulas 
(\ref{ap1}) given in appendix \ref{appen3}.

The structures ({\it 1-16}) of (\ref{eq34c}) are forming the basis for 
the relativistic nucleon wave function. The coefficients in the 
decomposition of the wave function in terms of this basis are the 
scalar functions. They depend on the quark momenta $\vec{q}_{1-3}$ and 
on $\vec{n}$ and are found from dynamics.

In the nonrelativistic limit the wave function does not depend on the 
orientation of the light-front plane, hence the contribution of the 
components depending on $\vec{n}$ disappears (as well as the 
$\vec{n}$-dependence of the scalar functions). The wave function in 
this limit contains eight structures of eqs. (\ref{eq34c}), namely, 
({\it 1-5}), ({\it 8}) and two $\vec{n}$-independent structures of 
({\it 11-16}).  This just corresponds to the fact that the momentum 
conservation in the nonrelativistic limit obtains the form:  
$\vec{p}=\vec{k}_1+\vec{k}_2+\vec{k}_3$, and in the system of the rest 
$\vec{p}=0$ we have only two independent three-vectors. Note that this 
reduction is a peculiarity of the two- and three-body systems.  In a 
system of $n$ particles with $n > 3$ the nonrelativistic limit does not 
decrease the number of components.  For example, the wave function of 
$^4$He contains 16 components both in relativistic and in 
nonrelativistic cases.

The basis (\ref{eq34c}) is not yet irreducible respective to the 
permutation group.  The corresponding irreducible basis will be 
constructed below.

We also list all the spin structures in the four-dimensional form:
\begin{eqnarray}
({\it 1}) &&  [\bar{u}(k_1) U_c\bar{u}(k_2)][\bar{u}(k_3)\gamma_5 u(p)]
\nonumber\\
({\it 2}) && [\bar{u}(k_1)\gamma_5 U_c\bar{u}(k_2)] [\bar{u}(k_3)u(p)]
\nonumber\\
({\it 3}) && [\bar{u}(k_1)\hat{p}\gamma_5 U_c\bar{u}(k_2)] 
[\bar{u}(k_3)u(p)] \nonumber\\ 
({\it 4}) && [\bar{u}(k_1)\hat{p} 
U_c\bar{u}(k_2)] [\bar{u}(k_3)\gamma_5 u(p)] 
\nonumber\\ 
({\it 5}) && [\bar{u}(k_1)\gamma^{\mu} U_c\bar{u}(k_2)] 
[\bar{u}(k_3)\gamma_{\mu}\gamma_5u(p)]
\nonumber\\
({\it 6}) && [\bar{u}(k_1)\gamma^{\mu}\gamma_5 U_c\bar{u}(k_2)]
[\bar{u}(k_3)\gamma_{\mu} u(p)]
\nonumber\\
({\it 7}) && [\bar{u}(k_1)\sigma^{\mu\nu}p_{\nu} U_c\bar{u}(k_2)]
[\bar{u}(k_3)\gamma_{\mu}\gamma_5u(p)]
\nonumber\\
({\it 8}) && [\bar{u}(k_1)\sigma^{\mu\nu} U_c\bar{u}(k_2)]
[\bar{u}(k_3)\sigma_{\mu\nu}\gamma_5u(p)]
\label{eq86d}\\
&&
\nonumber\\ 
({\it 9}) && C_{ps} [\bar{u}(k_1) U_c\bar{u}(k_2)][\bar{u}(k_3) u(p)]
\nonumber\\
({\it 10}) &&C_{ps} [\bar{u}(k_1)\gamma_5 U_c\bar{u}(k_2)] 
[\bar{u}(k_3)\gamma_5 u(p)]
\nonumber\\
({\it 11}) &&C_{ps} [\bar{u}(k_1)\hat{p} U_c\bar{u}(k_2)] 
[\bar{u}(k_3)u(p)] 
\nonumber\\ 
({\it 12}) &&C_{ps} [\bar{u}(k_1)\hat{p}\gamma_5 U_c\bar{u}(k_2)] 
[\bar{u}(k_3)\gamma_5 u(p)] 
\nonumber\\ 
({\it 13}) &&C_{ps} [\bar{u}(k_1)\gamma^{\mu} U_c\bar{u}(k_2)] 
[\bar{u}(k_3)\gamma_{\mu}u(p)]
\nonumber\\
({\it 14}) &&C_{ps} [\bar{u}(k_1)\gamma^{\mu}\gamma_5 U_c\bar{u}(k_2)]
[\bar{u}(k_3)\gamma_{\mu}\gamma_5 u(p)]
\nonumber\\
({\it 15}) &&C_{ps} [\bar{u}(k_1)\sigma^{\mu\nu}p_{\nu} U_c\bar{u}(k_2)]
[\bar{u}(k_3)\gamma_{\mu}u(p)]
\nonumber\\
({\it 16}) &&C_{ps} [\bar{u}(k_1)\sigma^{\mu\nu} U_c\bar{u}(k_2)]
[\bar{u}(k_3)\sigma_{\mu\nu}u(p)]
\label{eq86f}
\end{eqnarray}
where $C_{ps}$ is the pseudoscalar:
\begin{equation}\label{eq86e}
C_{ps}=\epsilon^{\mu\nu\rho\gamma}k_{1\mu}k_{2\nu}k_{3\rho}p_{\gamma}.
\end{equation}
By means of eq. (\ref{eq29a}) $C_{ps}$ can be 
transformed as:
$
C_{ps}\propto \epsilon^{\mu\nu\rho\gamma}
k_{1\mu}k_{2\nu}\omega_{\rho}{\cal P}_{\gamma}.
$
In the system of reference where $\vec{\cal P}=0$ it turns into 
$\vec{n}\cd[\vec{q}_1\times \vec{q}_2] $.

The structures ({\it 1-8}) of (\ref{eq86d}) exactly coincide with ones 
given in the paper \cite{dz88}. The structures ({\it 9-16}) of 
(\ref{eq86f}) are obtained from ({\it 1-8}) by deleting or adding 
$\gamma_5$ and multiplying by $C_{ps}$, that does not change their 
parity.  They disappear on the energy shell $p=k_1+k_2+k_3$, since in 
this case $C_{ps}=0$.

Any other structure is expressed through these ones. For example, 
instead of the structures ({\it 11}),({\it 12}) and ({\it 15}) one can 
take the structures obtained from ({\it 3}),({\it 4}) and ({\it 7}) by 
the replacement $p\rightarrow k_3$, and reduce by this way the number 
of structures containing the factor $C_{ps}$.

There is no any one-to-one correspondence between a given 
three-dimensional structures (\ref{eq34c}) and one of the  
four-dimensional structures (\ref{eq86d}-\ref{eq86f}).  A given 
structure from (\ref{eq86d}-\ref{eq86f}) is expressed through a linear 
combination of (\ref{eq34c}).  The increase of the number of components 
from 8 to 16 in comparison to the paper \cite{dz88} {\it is not only} 
due to the $\vec{n}$-depending structures. The latters are already 
implicitly included in the ``old" structures eq. (\ref{eq86d}).  For 
example, the structure ({\it 1}) of eq.  (\ref{eq86d}) generates the 
$\vec{n}$-depending  three-dimensional structure of the type 
$(\vec{\sigma}_{12}\cd \vec{q}_1) (\vec{\sigma}_{3N}\cd \vec{n})$, 
since at $\vec{\cal P}=0$ we get in the argument of the spinor $u(p)$:  
$\vec{p}=-\vec{\omega}\tau\propto \vec{n}$. 

The $\vec{n}$-depending components may be important. The example is 
given by the deuteron wave function \cite{ck-deut}, where 
$\vec{n}$-depending component, so called $f_5$, dominates starting with 
$q=0.5$ GeV/c.

We emphasize that the extra structures ({\it 9-16}) in (\ref{eq86f}) 
are not a peculiarity of the covariant formulation of LFD. The 
four-vector $\omega$ even does not enter in (\ref{eq86f}).  Therefore 
another eight structures of the nucleon wave function should appear in 
any light-front approach.

In the following we will transform the basis (\ref{eq34c}) 
to the form providing the irreducible representations of the 
permutation group and represent it in terms of the bi-spinors. 

\section{The permutation group in the three-body case}\label{pg}
\subsection{The irreducible representations}\label{irred}
We briefly remind the general construction of the irreducible 
representations of the permutation group in the three-body case.  

By permutations of the arguments of the function $\psi(123)$ we get six 
functions. Define their six linear combinations corresponding to the 
Young diagrams.

Fully symmetric:
\begin{equation}\label{eq1}
\psi_S=
\begin{tabular}{|c|c|c|}
\hline
1&2&3\\
\hline
\end{tabular}
=\psi(123)+\psi(213)+\psi(231)+\psi(321)+\psi(312)+\psi(132).
\end{equation}

Fully antisymmetric:
\begin{equation}\label{eq2}
\psi_A=
\begin{tabular}{|c|}
\hline
1\\
\hline
2\\
\hline
3\\
\hline
\end{tabular}
=\psi(123)-\psi(213)+\psi(231)-\psi(321)+\psi(312)-\psi(132).
\end{equation}

Mixed symmetry doublet:
\begin{equation}\label{eq3}
\left\{
\begin{array}{l}
\psi_1=
\begin{array}{cc}
\begin{tabular}{|c|}
\hline
1\\
\hline
\end{tabular}
&
\!\!\!\!\!
\begin{tabular}{c|}
\hline
2\\
\hline
\end{tabular}\\
\begin{tabular}{|c|}
3\\
\hline
\end{tabular}
&
\end{array}
=\psi(123)+\psi(213)-\psi(321)-\psi(231)\\
\\
\psi_2=
\begin{array}{cc}
\begin{tabular}{|c|}
\hline
2\\
\hline
\end{tabular}
&
\!\!\!\!\!
\begin{tabular}{c|}
\hline
1\\
\hline
\end{tabular}\\
\begin{tabular}{|c|}
3\\
\hline
\end{tabular}
&
\end{array}
=\psi(123)+\psi(213)-\psi(132)-\psi(312).
\end{array}\right.
\end{equation}

Another mixed symmetry doublet:
\begin{equation}\label{eq4}
\left\{
\begin{array}{l}
\psi''_1=
\begin{array}{cc}
\begin{tabular}{|c|}
\hline
1\\
\hline
\end{tabular}
&
\!\!\!\!\!
\begin{tabular}{c|}
\hline
3\\
\hline
\end{tabular}\\
\begin{tabular}{|c|}
2\\
\hline
\end{tabular}
&
\end{array}
=\psi(123)+\psi(321)-\psi(213)-\psi(312)\\
\\
\psi''_2=
\begin{array}{cc}
\begin{tabular}{|c|}
\hline
3\\
\hline
\end{tabular}
&
\!\!\!\!\!
\begin{tabular}{c|}
\hline
1\\
\hline
\end{tabular}\\
\begin{tabular}{|c|}
2\\
\hline
\end{tabular}
&
\end{array}
=\psi(123)+\psi(321)-\psi(132)-\psi(231).
\end{array}\right.
\end{equation}
Relative to the permutations $P_{ij}$ of the particles $ij$ the mixed 
symmetry doublets are transformed as follows:
\begin{eqnarray}\label{eq5}
&P_{12}\left(
\begin{array}{l}
\psi_1\\
\psi_2
\end{array}
\right)
=
\left(
\begin{array}{l}
\psi_2\\
\psi_1
\end{array}
\right),\;
P_{13}\left(
\begin{array}{l}
\psi_1\\
\psi_2
\end{array}
\right)
=
\left(
\begin{array}{l}
-\psi_1\\
-\psi_1+\psi_2
\end{array}
\right),\;
P_{23}\left(
\begin{array}{l}
\psi_1\\
\psi_2
\end{array}
\right)
=
\left(
\begin{array}{l}
\psi_1-\psi_2\\
-\psi_2
\end{array}
\right)\;\;&\\
&P_{12}\left(
\begin{array}{l}
\psi''_1\\
\psi''_2
\end{array}
\right)
=
\left(
\begin{array}{l}
-\psi''_1\\
-\psi''_1+\psi'_2
\end{array}
\right),\;
P_{13}\left(
\begin{array}{l}
\psi''_1\\
\psi''_2
\end{array}
\right)
=
\left(
\begin{array}{l}
\psi''_2\\
\psi''_1
\end{array}
\right),\;
P_{23}\left(
\begin{array}{l}
\psi''_1\\
\psi''_2
\end{array}
\right)
=
\left(
\begin{array}{l}
\psi''_1-\psi''_2\\
-\psi''_2
\end{array}
\right)\;\;&
\label{eq6}
\end{eqnarray}
Instead of (\ref{eq4}) we introduce another pair of the mixed symmetry 
functions:
\begin{equation}\label{eq6a}
\left\{
\begin{array}{lr}
\psi'_1=\psi''_2-\psi''_1
=&\psi(213)+\psi(312)-\psi(132)-\psi(231),\\
\psi'_2=\psi''_2\quad
=&\psi(123)+\psi(321)-\psi(132)-\psi(231).
\end{array}\right.
\end{equation}
Relative to permutations they are transformed identically to eq.  
(\ref{eq5}). Hence, we can consider the different mixed symmetry 
doublets with identical transformation properties (\ref{eq5}).

\subsection{The products of two irreducible representations}
\label{irreduc}
Below we will construct the spin-isospin functions $\phi$ with a given 
symmetry from the products of irreducible representations formed by the 
spin functions $\varphi$ with a given symmetry and by the mixed 
symmetry isospin functions $\chi$.  As well known, there are no any 
fully symmetric and antisymmetric isospin functions $\chi_S,\chi_A$ 
(see below sect. \ref{isospin}).  Then from the 
spin-isospin functions $\phi$ and the invariant functions $\psi$, 
depending on momenta, we will construct the fully symmetric nucleon 
wave functions ${\mit\Psi}$ (implicitly multiplied by the antisymmetric 
color singlet wave function).

We start with most general formula containing all possible 
products of $\varphi$'s and $\chi$'s:
\begin{equation}\label{eq7}
\left\{
\begin{array}{l}
\phi_1 =a_1\varphi_1\chi_1+b_1\varphi_2\chi_2+c_1\varphi_1\chi_2 
+d_1\varphi_2\chi_1,\\
\phi_2 =a_2\varphi_1\chi_1+b_2\varphi_2\chi_2+c_2\varphi_1\chi_2 
+d_2\varphi_2\chi_1,
\end{array}\right.
\end{equation}
where $\varphi_{1,2}$ and $\chi_{1,2}$ satisfy (\ref{eq5}). We also 
consider the similar formulas  with $\varphi$ fully symmetric and 
antisymmetric.  We find the coefficients in (\ref{eq7}) from the 
conditions that these $\phi_1,\phi_2$ were fully symmetric, 
antisymmetric or also satisfy (\ref{eq5}). 

For the fully symmetric function $\phi_S$ we find:
\begin{equation}
\label{eq9cc}
\phi_S=2(\varphi_1\chi_1+\varphi_2\chi_2)
-\varphi_1\chi_2-\varphi_2\chi_1.
\end{equation}

The fully antisymmetric function reads:
\begin{equation}
\label{eq13c}
\phi_A=\varphi_1\chi_2-\varphi_2\chi_1.
\end{equation}
For the mixed symmetry functions we obtain:
\begin{equation}\label{eq14c}
\left\{
\begin{array}{l}
\phi_1=\varphi_S\chi_1\\
\phi_2=\varphi_S\chi_2
\end{array}
\right. (a) \quad
\left\{
\begin{array}{l}
\phi_1=\varphi_A(\chi_1-2\chi_2)\\
\phi_2=\varphi_A(2\chi_1-\chi_2)
\end{array}
\right. (b)\quad
\left\{
\begin{array}{l}
\phi_1=\;\;\,\varphi_1(\chi_1-\chi_2)-\varphi_2\chi_1\\
\phi_2=-\varphi_1\chi_2+\varphi_2(\chi_2-\chi_1)
\end{array}
\right. (c)
\end{equation}

By this way we find for the total fully symmetric 
nucleon wave function:
\begin{eqnarray} 
&&{\mit\Psi}_S=\psi_S\phi_S,\quad
\label{eq8}\\
&&{\mit\Psi}_S=\psi_A\phi_A
\label{eq21a}\\
&&{\mit\Psi}_S=2(\psi_1\phi_1+\psi_2\phi_2)
-\psi_1\phi_2-\psi_2\phi_1
\label{eq9}
\end{eqnarray}
One should substitute in (\ref{eq9}) the different mixed symmetry 
doublets $\phi_{1,2}$ from (\ref{eq14c}).

\section{The components independent of momenta}
\label{indep}
\subsection{The spin-isospin functions with a given symmetry}
\label{isospin}
To establish the permutation properties, we put $\vec{\cal P}=0$ and 
denote the spin functions ({\it 1-2}) of eq. (\ref{eq34c}), 
corresponding to the spins of pairs 0 and 1, as:  
\begin{equation}\label{eq19} 
{\varphi^{(0)}}\equiv {\varphi}^{(0)}(123)=
(w_1^\dagger\sigma_yw_2^*)(w_3^\dagger w_N),
\quad
{\varphi}^{(1)}\equiv {\varphi}^{(1)}(123)=
(w_1^\dagger\vec{\sigma}\sigma_yw_2^*) 
(w_3^\dagger\vec{\sigma}w_N).
\end{equation} 
We keep the order of the quark spinors 123.  Using the Fierz 
identities, given in appendix \ref{Fierz}, one can establish the 
following permutation properties:  
\begin{eqnarray}\label{eq20} 
{\varphi}^{(0)}(123)=-{\varphi}^{(0)}(213)={\varphi}^{(0)},
&&{\varphi}^{(0)}(321)=-{\varphi}^{(0)}(231)=
\frac{1}{2}{\varphi}^{(0)}+\frac{1}{2}{\varphi}^{(1)},
\nonumber\\
&&{\varphi}^{(0)}(132)=-{\varphi}^{(0)}(312)=
\frac{1}{2}{\varphi}^{(0)}-\frac{1}{2}{\varphi}^{(1)};
\nonumber\\
{\varphi}^{(1)}(123)={\varphi}^{(1)}(213)={\varphi}^{(1)},
&&{\varphi}^{(1)}(321)={\varphi}^{(1)}(231)=
\quad\frac{3}{2}{\varphi}^{(0)}-\frac{1}{2}{\varphi}^{(1)},
\nonumber\\
&&{\varphi}^{(1)}(132)={\varphi}^{(1)}(312)=
-\frac{3}{2}{\varphi}^{(0)}-\frac{1}{2}{\varphi}^{(1)}.
\end{eqnarray}
We take as $\psi(123)$ in the formulas of sect. \ref{irred} the spin 
wave function $\varphi(123)$ of the general form:
\begin{equation}\label{eq22}
\varphi(123)=a{\varphi}^{(1)}(123)+b{\varphi}^{(0)}(123).
\end{equation}
Substituting (\ref{eq22}) in eqs. (\ref{eq1}), (\ref{eq2}) and using 
(\ref{eq20}) we reproduce the well known result:
$$
\varphi_S=0,\quad
\varphi_A=0.
$$
Substituting (\ref{eq22}) in eqs. (\ref{eq3}),
for the functions with mixed symmetry we find:
$$
\left\{
\begin{array}{l}
\varphi_1\propto 3a(-{\varphi}^{(0)}+{\varphi}^{(1)}),\\
\varphi_2\propto 3a(\quad {\varphi}^{(0)}+{\varphi}^{(1)}).
\end{array}
\right.
$$
The second mixed symmetry doublet $\varphi'_1,\varphi'_2$, eq. 
(\ref{eq6a}), coincides (up to a factor) with $\varphi_1,\varphi_2$.

Denote:
\begin{equation}\label{eq13}
\left\{
\begin{array}{l}
\varphi_1=-{\varphi}^{(0)}+{\varphi}^{(1)}=
-1+\vec{\sigma}_{12}\cd\vec{\sigma}_{3N},\\
\varphi_2=\quad\! {\varphi}^{(0)}+{\varphi}^{(1)}=\quad\! 
1+\vec{\sigma}_{12}\cd\vec{\sigma}_{3N}.  
\end{array}\right.
\end{equation} 
These functions are forming the mixed symmetry doublet of the type 
(\ref{eq3}). The normalization factor will be determined for full wave 
function.

By this way, after permutations we have reduced the spin wave function
to the form (\ref{eq13}) with the ``canonical" order of spinors 
$(12)(3N)$, when 1 is coupled with 2 and 3 with $N$. In principle, this 
is not obligatory, but is very convenient.  We will see below in sect. 
\ref{appl} that in the form factor calculations this results in the 
product of two traces, each of them containing approximately half of 
the Dirac matrices involved, instead of the trace of full product.  
This simplifies calculations considerably.

Similarly to (\ref{eq13}) we introduce the isospin functions:
\begin{equation}\label{eq23}
\left\{
\begin{array}{l}
\chi_1=-\chi^{(0)}+\chi^{(1)}=
-1+\vec{\tau}_{12}\cd\vec{\tau}_{3N},\\
\chi_2=\quad\!\chi^{(0)}+\chi^{(1)}=
\quad\! 1+\vec{\tau}_{12}\cd\vec{\tau}_{3N},
\end{array}\right.
\end{equation}
where 
\begin{equation}\label{eq24} 
1=({\xi}_1^\dagger\tau_y 
{\xi}_2^*)({\xi}_3^\dagger {\xi}_N)\equiv \chi^{(0)},\quad 
\vec{\tau}_{12}\cd\vec{\tau}_{3N}= ({\xi}_1^\dagger\vec{\tau}\tau_y 
{\xi}_2^*)\cd ({\xi}_3^\dagger\vec{\tau}{\xi}_N)\equiv \chi^{(1)}, 
\end{equation} 
$\xi$'s are the isospin spinors. These functions satisfy the 
symmetry properties (\ref{eq20}).  Like in the case of 
spin, $\chi_S=\chi_A=0$.

From the above mixed symmetry spin and isospin doublets we can 
construct the spin-isospin functions with given symmetry.

Substituting (\ref{eq13}) and (\ref{eq23}) into the equation 
(\ref{eq9cc}), we get the symmetric spin-isospin function. Multiplying 
it by the symmetric $\psi_S$, we find the total symmetric nucleon wave 
function:
\begin{equation}\label{eq27} 
{\mit\Psi}_S=\frac{1}{\sqrt{72}}
\psi_S[3+(\vec{\sigma}_{12}\cd\vec{\sigma}_{3N})
(\vec{\tau}_{12}\cd\vec{\tau}_{3N})].
\end{equation}
It is given in many papers (compare, for example, with \cite{cc91}).  
The invariant symmetric function $\psi_S$ depends on the relative 
momenta and on $\vec{n}$: $\psi_S=\psi_S(\vec{q}_1,\vec{q}_2, 
\vec{q}_3, \vec{n})$. The normalization factor will be explained in the 
next section. 

Substituting (\ref{eq13}) and (\ref{eq23}) into the equation 
(\ref{eq13c}), we get for the antisymmetric spin-isospin function.
Multiplying it by antisymmetric $\psi_A$, we find:
\begin{equation}\label{eq27a}
{\mit\Psi}_S=\frac{1}{\sqrt{24}}\psi_A[\vec{\sigma}_{12}\cd\vec{\sigma}_{3N} 
-\vec{\tau}_{12}\cd\vec{\tau}_{3N} ]
\end{equation} 
Substituting (\ref{eq13}) and (\ref{eq23}) into (\ref{eq14c}), we 
obtain the mixed symmetry functions: 
\begin{equation}\label{eq26a}
\left\{
\begin{array}{l}
\phi_1=
3-\vec{\tau}_{12}\cd\vec{\tau}_{3N} -\vec{\sigma}_{12}\cd\vec{\sigma}_{3N}
-(\vec{\sigma}_{12}\cd\vec{\sigma}_{3N}) 
(\vec{\tau}_{12}\cd\vec{\tau}_{3N})\\
\phi_2=3+\vec{\tau}_{12}\cd\vec{\tau}_{3N} 
+\vec{\sigma}_{12}\cd\vec{\sigma}_{3N} 
-(\vec{\sigma}_{12}\cd\vec{\sigma}_{3N}) 
(\vec{\tau}_{12}\cd\vec{\tau}_{3N}) 
\end{array} \right.  \end{equation} 
Substituting eqs. (\ref{eq26a}) into (\ref{eq9}) we get:  
\begin{eqnarray}\label{eq28} 
{\mit\Psi}_S&=&\frac{1}{\sqrt{288}}\psi_1
[3-3\vec{\tau}_{12}\cd\vec{\tau}_{3N} -3\vec{\sigma}_{12}\cd\vec{\sigma}_{3N}
-(\vec{\sigma}_{12}\cd\vec{\sigma}_{3N}) (\vec{\tau}_{12}\cd\vec{\tau}_{3N})]
\nonumber\\
&+&\frac{1}{\sqrt{288}}\psi_2
[3+3\vec{\tau}_{12}\cd\vec{\tau}_{3N} +3\vec{\sigma}_{12}\cd\vec{\sigma}_{3N}
-(\vec{\sigma}_{12}\cd\vec{\sigma}_{3N}) (\vec{\tau}_{12}\cd\vec{\tau}_{3N})]
\end{eqnarray}

\subsection{Normalization}\label{norm1}
To calculate the normalization, we use the equalities
\begin{eqnarray}\label{eq1n}
\varphi^{\dagger (0)}\varphi^{(0)}&=&\frac{1}{2}\sum_{\sigma, \sigma_1, 
\sigma_2, \sigma_3} [(w_2^{\sigma_2}\sigma_y w_1^{\sigma_1}) 
(w_1^{\dagger\sigma_1}\sigma_y w_2^{*\sigma_2})]\;
[(w^{\dagger\sigma}_N w_3^{\sigma_3})
(w_3^{\dagger\sigma_3} w^{\sigma}_N)]
\nonumber\\
&=&\frac{1}{2}Tr[\sigma_y^2]\;Tr[1]=2,
\nonumber\\
\varphi^{\dagger (1)}\varphi^{(1)}&=&\frac{1}{2}\sum_{\sigma, \sigma_1, 
\sigma_2, \sigma_3} [(w_2^{\sigma_2}\sigma_i\sigma_y w_1^{\sigma_1}) 
(w_1^{\dagger\sigma_1}\sigma_y\sigma_j w_2^{*\sigma_2})]\;
[(w^{\dagger\sigma}_N\sigma_i w_3^{\sigma_3})
(w_3^{\dagger\sigma_3}\sigma_j w^{\sigma}_N)]
\nonumber\\
&=&\frac{1}{2}Tr[\sigma_j\sigma_i]\;Tr[\sigma_i\sigma_j]=6,
\nonumber\\
\varphi^{\dagger (1)}\varphi^{(0)}&=&\varphi^{\dagger 
(0)}\varphi^{(1)}=0, 
\end{eqnarray} 
and similarly for the isospin 
functions ${\chi}^{(0)},{\chi}^{(1)}$. The factor 
$\frac{\displaystyle{1}} {\displaystyle{2}}$ in these formulas appears 
due to averaging over the nucleon spin projection $\sigma$.  Then for 
the function (\ref{eq27}) we find:  \begin{equation}\label{eq3n} 
\overline{{\mit\Psi}^\dagger_S{\mit\Psi}_S}= 
\frac{1}{72}(3^2\times 2\times 2+6\times 6)\psi_S^2 =\psi_S^2.
\end{equation}
The factor $2\times 2$ in the first item comes from the product of the 
spin and isospin functions $\varphi^{(0)},{\chi}^{(0)}$, the factor 
$6\times 6$ in the second item comes from  the product of 
$\varphi^{(1)},{\chi}^{(1)}$. 

For the function (\ref{eq27a}) with antisymmetric $\psi_A$ we get:
\begin{equation}\label{eq4n}
\overline{{\mit\Psi}^\dagger_S{\mit\Psi}_S}= 
\frac{1}{24}(6\times 2 +2\times 6)\psi_A^2 = \psi_A^2,
\end{equation}
and for the mixed symmetry $\psi_{1,2}$, eq. (\ref{eq28}):
\begin{equation}\label{eq5n}
\overline{{\mit\Psi}^\dagger_S{\mit\Psi}_S}= (\psi_1^2+\psi_2^2-\psi_1\psi_2).
\end{equation}
Provided the function $\psi_S$ dominates, it is normalized as
\begin{equation}\label{eq6n}
\int \psi_S^2D=1.
\end{equation}
If $\psi_S$ does not dominate, the integral (\ref{eq6n}) is not equal 
to 1, but gives the contribution of $\psi_S$ into full normalization 
integral. The contributions of symmetric and mixed symmetry states are 
given by the similar integrals from $\psi_A^2$ and
$(\psi_1^2+\psi_2^2-\psi_1\psi_2)$.

In eq. (\ref{eq6n}) we introduced for shortness the notation for the 
integration volume \cite{cdkm,km96}: 
\begin{eqnarray}\label{b1}
D&\equiv&
(2\pi)^3\delta^{(3)}(\sum_{i=1}^3\vec{q}_i)2 
(\sum_{i=1}^3\varepsilon_{q_i}) \prod_{i=1}^3{d^3q_i\over (2\pi)^3 
2\varepsilon_{q_i}}.
\nonumber\\
&=&(2\pi)^3\delta^{(2)}(\sum_{i=1}^3 \vec{R}_{i\perp}) \delta(\sum_{i=1}^3 
x_i-1) 2\prod_{i=1}^3\frac{d^2R_{i\perp}}{(2\pi)^3}\frac{dx_i}{2x_i}
\end{eqnarray}  
The last line of (\ref{b1}) is written in the variables which are 
similar to the well known infinite momentum frame  variables.  Namely, 
$x_i$ is defined as $x_i=\omega\cd k_i/\omega\cd p$.  Then we introduce 
the four-vectors: $R_i=k_i-x_ip$.  They satisfy the condition 
$R_i\cd\omega=0$ and can be represented as $R_i=(R_{i0}, 
\vec{R}_{i\perp},\vec{R}_{i\parallel})$, where  $\vec{R}_{i\parallel}$ 
is parallel to $\vec{\omega}$ and $\vec{R}_{i\perp} \cd 
\vec{\omega}=0$. So, $\vec{R}_{i\perp}$ are the two-dimensional 
vectors such that $\vec{R}_{i\perp}^2=-R_i^2$. We thus have the 
following relations: 
$$\vec{R}_{1\perp}+\vec{R}_{2\perp}+\vec{R}_{3\perp}=0,\quad
x_1+x_2+x_3=1,$$
like for the infinite momentum frame  variables.

For convenience, we can express $\psi_S$ in (\ref{eq27}) through 
another  scalar function $\varphi_0$:
\begin{equation}\label{wf6} 
\psi_S=\frac{2}{\sqrt{3}}Nm \varphi_0\ ,
\end{equation}
with the normalization
\begin{equation}\label{wf6p}
\int\varphi_0^2(2\pi)^3\delta^{(3)}(\vec{q}_1+\vec{q}_2+\vec{q}_3)
\prod_{i=1}^3\frac{d^3q_i}{(2\pi)^3}=1\ .
\end{equation}
The dimensionless normalization constant $N$ in (\ref{wf6}) is found 
from the condition obtained by substituting (\ref{wf6}) into 
(\ref{eq6n}):
\begin{equation}\label{wf6pp}
\frac{4}{3}N^2 m^2
\int \varphi_0^2D=1.
\end{equation}  
In the nonrelativistic limit the factor $D$ turns into the integration 
volume of eq. (\ref{wf6p}) multiplied by $3/(4m^2)$. Therefore $N$ tends 
to 1. 

For $\varphi_0$ we can take, for example, the harmonic oscillator 
model:  
\begin{equation}\label{wf6a}
\varphi_0=\frac{2^3\pi^{3/2}3^{3/4}}{\alpha^3}
\exp\left(-\frac{\vec{q}_1\,^2 
+\vec{q}_2\,^2+\vec{q}_3\,^2}{2\alpha^2}\right)\ .
\end{equation}

\subsection{Four-dimensional representation}
In the section \ref{so} the spin operators 
$\vec{\sigma}_{12},\vec{\sigma}_{3N}$ were constructed through  the 
spinors $u_{\cal P}$ in the special representation (\ref{sp7}). In this 
section we represent the structures given above through the usual 
spinors $u$. This allows to use in calculations the standard trace 
techniques.  Similar representation for the components (\ref{eq86d}) on 
the energy shell was found in \cite{bkw}.

Since the spinors in two representations are related with each other by 
eq. (\ref{sp7}), the standard representation is obtained by replacing 
$u_{\cal P}$ by $u$. They coincide with each other in the reference 
frame where $\vec{\cal P}$. So, from (\ref{eq80n}) we find that:
\begin{equation}\label{eq83}
\vec{\sigma}_{12}\cd\vec{\sigma}_{3N}\rightarrow 
-c_1c_2c_3c_N[\bar{u}(k_1)\Pi_+\gamma^{\mu}\Pi_-U_c\bar{u}(k_2)]
[\bar{u}(k_3)\Pi_+\gamma_{\mu}\gamma_5\Pi_+ u(p)].
\end{equation}
The scalar product is defined as $a\cd b=a_{\mu}b^{\mu}=a_0b^0 
-\vec{a}\cd\vec{b}$. At $\vec{\cal P}=0$ we have $\Pi_+\gamma^0\Pi_- = 
\Pi_+\gamma^0\gamma_5\Pi_+ =0$, and we see that r.h.-side of 
(\ref{eq83}) indeed coincides with 
$\vec{\sigma}_{12}\cd\vec{\sigma}_{3N}$.

Similarly we obtain for the unit operators:
\begin{equation}\label{eq82}
1_{12}=(w_1^{\dagger}\sigma_yw_2^*)\rightarrow 
c_1c_2\bar{u}(k_1)\Pi_+\gamma_5\Pi_- U_c\bar{u}(k_2),\quad
1_{3N}=(w_3^{\dagger}w_N)\rightarrow 
c_3c_N\bar{u}(k_3)\Pi_+\Pi_+u(p).
\end{equation}
In eqs. (\ref{eq82}) we can substitute 
$\Pi_+\gamma_5\Pi_-=\Pi_+\gamma_5$, $\Pi_+\Pi_+=\Pi_+$. 

Using the above formulas we represent in the four-dimensional form the 
functions with the spin structures independent of momenta. The 
symmetric structure (\ref{eq27}) is represented as:
\begin{eqnarray}\label{eq86p} 
{\mit\Psi}_S&=&\frac{\psi_S}{\sqrt{72}} 
c_1c_2c_3c_N\{3[\bar{u}(k_1)\Pi_+\gamma_5  U_c\bar{u}(k_2)] 
[\bar{u}(k_3)\Pi_+ u(p)] 
\nonumber\\ 
&&-[\bar{u}(k_1)\Pi_+\gamma^{\mu}\Pi_- U_c\bar{u}(k_2)]
[\bar{u}(k_3)\Pi_+\gamma_{\mu}\gamma_5 \Pi_+ u(p)]
(\vec{\tau}_{12}\cd\vec{\tau}_{3N})\}
\nonumber\\ 
&\equiv& {\mit \Psi}_0+{\mit \Psi}_1,
\end{eqnarray}
where $c_i$ are defined by (\ref{eq81a}).  For applications given below 
we denote two items in (\ref{eq86p}) as ${\mit \Psi}_0$ and ${\mit 
\Psi}_1$.

The antisymmetric structure (\ref{eq27a}) reads:
\begin{eqnarray}\label{eq87} 
{\mit\Psi}_S&=&-\frac{\psi_A}{\sqrt{24}} c_1c_2c_3c_N\{
[\bar{u}(k_1)\Pi_+\gamma^{\mu}\Pi_- U_c\bar{u}(k_2)]
[\bar{u}(k_3)\Pi_+\gamma_{\mu}\gamma_5 \Pi_+ u(p)]
\nonumber\\
&&+[\bar{u}(k_1)\Pi_+\gamma_5  U_c\bar{u}(k_2)]
[\bar{u}(k_3)\Pi_+ u(p)]
(\vec{\tau}_{12}\cd\vec{\tau}_{3N}) \}.
\end{eqnarray} 
The mixed symmetry function (\ref{eq28}) is given by:
\begin{eqnarray}\label{eq88} 
{\mit\Psi}_S&=&\frac{\psi_1}{\sqrt{288}} c_1c_2c_3c_N\{
3[\bar{u}(k_1)\Pi_+\gamma_5 U_c\bar{u}(k_2)]
[\bar{u}(k_3)\Pi_+ u(p)] (1-\vec{\tau}_{12}\cd\vec{\tau}_{3N})
   \nonumber\\
&&\qquad +[\bar{u}(k_1)\Pi_+\gamma^{\mu}\Pi_- U_c\bar{u}(k_2)]
[\bar{u}(k_3)\pi_+\gamma_{\mu}\gamma_5 \pi_+ u(p)]
(3+\vec{\tau}_{12}\cd\vec{\tau}_{3N})\}
   \nonumber\\
& +&\frac{\psi_2}{\sqrt{288}}c_1c_2c_3c_N\{
3[\bar{u}(k_1)\Pi_+\gamma_5 U_c\bar{u}(k_2)]
[\bar{u}(k_3)\Pi_+ u(p)] (1+\vec{\tau}_{12}\cd\vec{\tau}_{3N})
   \nonumber\\
&&\qquad -[\bar{u}(k_1)\Pi_+\gamma^{\mu}\Pi_-U_c\bar{u}(k_2)]
[\bar{u}(k_3)\Pi_+\gamma_{\mu}\gamma_5 \Pi_+ u(p)]
(3-\vec{\tau}_{12}\cd\vec{\tau}_{3N})\}.
\end{eqnarray}

The above expressions can be represented in the form of decomposition 
in terms of the spin structures (\ref{eq86d}-\ref{eq86f}). They become 
more lengthy, and we will not give them here. The expressions 
(\ref{eq86p}-\ref{eq88}) are enough for calculations. Note, however, 
that the ``old" structures (\ref{eq86d}) are not enough for this 
decomposition. For example, the decomposition of the symmetric function 
(\ref{eq86p}) contains the terms from (\ref{eq86f}), constructed by 
means of the pseudoscalar $C_{ps}$, eq. (\ref{eq86e}). 

We give also the four-dimensional form of the normalization in the 
example of symmetric wave function. Its contribution to the 
normalization integral has the form:
\begin{eqnarray}\label{eq86b}
&&\int \overline{{\mit\Psi}^{\dagger}_S{\mit\Psi}_S} D,
\quad \mbox{where} \quad
\overline{{\mit\Psi}^{\dagger}_S{\mit\Psi}_S}=
\frac{1}{2\cd 72}\psi_S^2c_1^2 c_2^2 c_3^2 c_N^2
\nonumber\\
&\times&
\{3^2 2Tr[\Pi_-\gamma_5\Pi_+ (\hat{k}_1+m)\Pi_+ \gamma_5\Pi_- 
(-\hat{k}_2+m)] Tr[\Pi_+(\hat{k}_3+m)\Pi_+(\hat{p}+M)] 
\nonumber\\ 
&+&6Tr[\Pi_-\gamma^{\nu}\Pi_+(\hat{k}_1+m)\Pi_+\gamma^{\mu}\Pi_-
(-\hat{k}_2+m)]
\nonumber\\
&\times&
Tr[\Pi_+\gamma_5\gamma_{\nu}\Pi_+(\hat{k}_3+m)\Pi_+\gamma_{\mu}
\gamma_5\Pi_+(\hat{p}+M)]\}.
\end{eqnarray}
This formula can be simplified, using the equalities 
$$
\Pi_+(\hat{k_1}+m)\Pi_+ = \Pi_+/c_1^2,\quad
\Pi_-(-\hat{k_2}+m)\Pi_- = \Pi_-/c_2^2,\quad \mbox{etc.}
$$
By this way we get:
\begin{eqnarray}\label{eq90a}
\overline{{\mit\Psi}^{\dagger}_S{\mit\Psi}_S}&=&
\psi_S^2\frac{1}{2\cd 72}\{3^2 2 Tr[\Pi_+] Tr[\Pi_+]
+6 Tr[\Pi_-\gamma^{\nu}\Pi_+\gamma^{\mu}]
Tr[\Pi_-\gamma_{\nu}\Pi_+\gamma_{\mu}]\}
\nonumber\\
&=&\psi_S^2\frac{1}{2\cd 72}\{3^2\times 2\times 2\times 2+
6\times 12\}=\psi_S^2,
\end{eqnarray}
that coincides with the normalization (\ref{eq3n}).

\section{The components depending on momenta}\label{depen}
\subsection{Symmetric tensors}\label{symm}
Let us construct now the depending on momenta spin components with the 
given symmetry. We start with the tensors (\ref{eq46}), symmetric in 
the indices $ij$. However, they have no any definite symmetry relative 
to the permutations of the particles.  We can construct the symmetric
relative to the permutations function:  
\begin{equation}\label{eq47} 
\label{eq47a}
f_S=T_{11}+T_{22}+T_{33}=2T_{11}+2T_{22}+T_{12},
\end{equation}
and two doublets with mixed symmetry:
\begin{equation}\label{eq48}
\left\{
\begin{array}{l}
f_1=T_{11}-T_{33}=-T_{22}-T_{12}\\
f_2=T_{22}-T_{33}=-T_{11}-T_{12}
\end{array}\right. (a)\quad
\left\{
\begin{array}{l}
f'_1=T_{1n}-T_{3n}=2T_{1n}+T_{2n}\\
f'_2=T_{2n}-T_{3n}=2T_{2n}+T_{1n}
\end{array}\right. (b)
\end{equation}
For shortness we omit the indices $ij$.  One can easily check that the 
functions (\ref{eq48}) indeed satisfy the permutation properties 
(\ref{eq5}).  It is impossible to construct any antisymmetric 
structure.  

We represent also the spin tensor $\sigma^i_{12}\sigma^j_{3N}$ in the 
form of symmetric in the indices $ij$ traceless tensor:
\begin{equation}\label{eq44}
S^{ij}=\sigma_{12}^i\sigma_{3N}^j
+\sigma_{12}^j\sigma_{3N}^i-
\frac{2}{3}\vec{\sigma}_{12}\cd\vec{\sigma}_{3N}\delta^{ij}.
\end{equation}
Using the Fierz identities given in appendix \ref{Fierz} we find that 
$S^{ij}$ is fully symmetric also relative to permutation of any 
pair of the particles 1,2,3. This reflects the fact that there is no 
any other traceless symmetric tensor made of 
$\vec{\sigma}_{12},\vec{\sigma}_{3N}$.

The depending on momenta spin structures 
$\varphi_S,\varphi_{1,2},\varphi'_{1,2}$ are obtained by the 
contracting of $f_S$, eq. (\ref{eq47}), and $f_{1,2},f'_{1,2}$, eq.  
(\ref{eq48}), with the spin operator $S$:
\begin{equation}\label{eq44a}
(\varphi_S,\varphi_{1,2},\varphi'_{1,2}) =
(f_S, f_{1,2},f'_{1,2})\times S
\end{equation}
This contraction does not change the symmetry of the functions $f$'s. 
Therefore the symmetry of $\varphi$'s coincides with the symmetry of 
$f$'s.

By the standard formulas of the section \ref{irreduc} we construct from 
$\varphi$ and from the isospin functions  $\chi$ the spin-isospin 
functions $\phi$  of all symmetries:  symmetric, antisymmetric and 
mixed symmetry. Then from these functions $\phi$  and from the 
invariant momentum functions $\psi$ of different symmetries, by the 
formulas (\ref{eq8}-\ref{eq9}), we construct the total symmetric wave 
function of the nucleon. 

For example, contracting $S$ with the functions $f_{1,2}$, 
eq. (\ref{eq48}), we get the mixed symmetry spin functions:
$$
\left\{
\begin{array}{l}
\varphi_1=T_{11}^{ij}S_{ij}-T_{33}^{ij}S_{ij}\\
\varphi_2=T_{22}^{ij}S_{ij}-T_{33}^{ij}S_{ij}
\end{array}\right.
$$
Then we substitute these functions $\varphi_{1,2}$ and the mixed 
symmetry isospin functions $\chi_{1,2}$  (\ref{eq23})  into  
(\ref{eq9cc}) and get the fully symmetric spin-isospin function 
$\phi_S$. Multiplied by $\psi_S$, eq. (\ref{eq8}), it gives the
total fully symmetric nucleon wave function:
\begin{eqnarray} 
{\mit\Psi}_S &=& \psi_S\left\{
\left[(\vec{q}_1\cd \vec{\sigma}_{12})(\vec{q}_1\cd\vec{\sigma}_{3N})
-\frac{1}{3}\vec{q}\,_1^2 (\vec{\sigma}_{12}\cd\vec{\sigma}_{3N})\right]
(\vec{\tau}_{12}\cd\vec{\tau}_{3N}-3)\right.
\nonumber\\
&+&\quad\quad
\left[(\vec{q}_2\cd \vec{\sigma}_{12})(\vec{q}_2\cd\vec{\sigma}_{3N})
-\frac{1}{3}\vec{q}\,_2^2 (\vec{\sigma}_{12}\cd\vec{\sigma}_{3N})\right]
(\vec{\tau}_{12}\cd\vec{\tau}_{3N}+3) 
\nonumber\\
&-&\quad\,
\left.
2\left[(\vec{q}_3\cd \vec{\sigma}_{12})(\vec{q}_3\cd\vec{\sigma}_{3N})
-\frac{1}{3}\vec{q}\,_3^2 (\vec{\sigma}_{12}\cd\vec{\sigma}_{3N})\right]
\vec{\tau}_{12}\cd\vec{\tau}_{3N}\right\} .
\label{eq50p1}
\end{eqnarray}
Similarly we find all other functions forming the basis of the nucleon 
wave function.  The full set of these functions is given in appendix 
\ref{appwf1}. 

\subsection{Antisymmetric tensors}
\label{asymm}
Now consider the tensors (\ref{eq60}), antisymmetric in
the indices $ij$.
Introduce the spin operators antisymmetric in $ij$:
\begin{equation}\label{eq50b}
\Gamma^{ij}(123)={\sigma}_{12}^i {\sigma}_{3N}^j-
{\sigma}_{12}^j {\sigma}_{3N}^i,\;
\Gamma_+^{ij}(123)=-i\epsilon_{ijk}({\sigma}_{12}^k+ {\sigma}_{3N}^k),\;
\Gamma_-^{ij}(123)=-i\epsilon_{ijk}({\sigma}_{12}^k- {\sigma}_{3N}^k).
\end{equation}
Using the Fierz identities, appendix \ref{Fierz}, one can establish the 
following properties relative to permutations of the particles:
\begin{equation}\label{eq53b}
\begin{array}{lll}
\Gamma(231)= \Gamma(321)=\Gamma_-(123),
& \Gamma(312)= \Gamma(132)=\Gamma_+(123),&\Gamma(213)=\Gamma(123),\\
\Gamma_+(213)=\Gamma_+(312)=\Gamma_-(123),
& \Gamma_+(231)=\Gamma_+(132)=\Gamma(123),
& \Gamma_+(321)=\Gamma_+(123),\\
\Gamma_-(213)=\Gamma_-(231)=\Gamma_+(123),
& \Gamma_-(321)=\Gamma_-(312)=\Gamma(123),
& \Gamma_-(132)=\Gamma_-(123).
\end{array}
\end{equation}
From three operators (\ref{eq50b}) we construct another three operators 
forming the symmetric representation of the permutation group 
$\Sigma_S$ and the mixed symmetry one $\Sigma_{1,2}$:
\begin{equation}\label{eq53da}
\Sigma_S = \Gamma + \Gamma_+ + \Gamma_- 
\end{equation}
\begin{equation}\label{eq53db}
\left\{ \begin{array}{l}
\Sigma_1 = \Gamma - \Gamma_-\\
\Sigma_2 = \Gamma - \Gamma_+
\end{array}\right.
\end{equation}
From the tensors (\ref{eq60}) we construct antisymmetric and the mixed 
symmetry functions:
\begin{equation}\label{eq61a} 
f_A=A_{12},
\end{equation}
\begin{equation}\label{eq61b} 
\left\{ 
\begin{array}{l}
f_1=A_{1n}-A_{3n}=2A_{1n}+A_{2n}\\
f_2=A_{2n}-A_{3n}=2A_{2n}+A_{1n}
\end{array}\right. 
\end{equation}
It is impossible to construct any symmetric function.

Using the formulas from sect. \ref{irreduc} we find the spin wave 
functions $\varphi$ with given symmetry. Namely:
\begin{equation}\label{as1}
\varphi_S=2(f_1\Sigma_1+ f_2\Sigma_2)
- f_1\Sigma_2- f_2\Sigma_1,\\
\end{equation}
\begin{equation}\label{as1a}
\begin{array}{lr}
\varphi_A= f_2\Sigma_1-f_1\Sigma_2 \qquad& (a)\\ 
\varphi_A= f_A\Sigma_S\qquad & (b)
\end{array}
\end{equation}
and
\begin{equation}\label{as2}
\left\{
\begin{array}{l}
\varphi_1=\Sigma_S f_1\\
\varphi_2=\Sigma_S f_2
\end{array}
\right. (a)\quad
\left\{
\begin{array}{l}
\varphi_1=f_A(\Sigma_1-2\Sigma_2)\\
\varphi_2=f_A(2\Sigma_1-\Sigma_2)
\end{array}
\right. (b)\quad
\begin{array}{l}
\end{array}\quad
\left\{
\begin{array}{l}
\varphi_1=(f_1-f_2)\Sigma_1- f_1\Sigma_2\\
\varphi_2=-f_2\Sigma_1+(f_2-f_1)\Sigma_2
\end{array}\right. (c)
\end{equation}
In the above formulas we imply the summation over the indices $ij$.

Then again by means of eqs. (\ref{eq9cc}-\ref{eq14c}) we construct from 
these $\varphi$'s and the isospin functions $\chi$'s the spin-isospin 
functions $\phi$ and then, by eqs.  (\ref{eq8}-\ref{eq9}), from 
$\phi$'s and the momentum functions $\psi$'s we obtain the total 
nucleon wave function ${\mit\Psi}_S$.  As an example, we give here a
wave function with the symmetric spin-isospin and the scalar  parts:  
\begin{equation}\label{eq63a1} 
{\mit\Psi}_S=\psi_S
\{(\vec{q}_1\cd\vec{\sigma}_{12})(\vec{q}_2\cd\vec{\sigma}_{3N})
-(\vec{q}_2\cd\vec{\sigma}_{12})(\vec{q}_1\cd\vec{\sigma}_{3N})
+i\vec{\sigma}_{12}\cd[\vec{q}_1\times\vec{q}_2]
-i\vec{\sigma}_{3N}\cd[\vec{q}_1\times\vec{q}_2]\;
\vec{\tau}_{12}\cd\vec{\tau}_{3N}\}
\end{equation}
The full set of the corresponding wave functions is given in appendix 
\ref{appwf2}.

\subsection{Four-dimensional representation}
The operators $\vec{\sigma}_{12}\cd\vec{\sigma}_{3N}$ and 1 are already 
represented through the Dirac matrices and the bi-spinors by the 
formulas (\ref{eq83}) and (\ref{eq82}).  In this section we give the 
four-dimensional form of other structures.

Any scalar products of $\vec{\sigma}_{1,2}$ with 
the vectors $\vec{q}_{1,2,3},\vec{n}$ can be represented similarly 
to (\ref{eq83}). For example:
\begin{eqnarray}\label{eq84}
&&(\vec{q}_2\cd\vec{\sigma}_{12})\rightarrow 
c_1c_2\bar{u}(k_1)\Pi_+\hat{k}_2\Pi_-U_c\bar{u}(k_2),\quad
(\vec{q}_1\cd\vec{\sigma}_{3N})\rightarrow 
c_3c_N\bar{u}(k_3)\Pi_+\hat{k}_1\gamma_5\Pi_+ u(p)
\nonumber\\
&&(\vec{n}\cd\vec{\sigma}_{12})\rightarrow
c_1c_2\frac{{\cal M}}{\omega\cd p}
\bar{u}(k_1)\Pi_+\hat{\omega}\Pi_-U_c\bar{u}(k_2),\quad
(\vec{n}\cd\vec{\sigma}_{3N})\rightarrow
c_3c_N\frac{{\cal M}}{\omega\cd p} 
\bar{u}(k_3)\Pi_+\hat{\omega}\gamma_5\Pi_+ u(p),
\nonumber\\
&&
\end{eqnarray}
etc. At $\vec{\cal P}=0,{\cal P}_0={\cal M}$ the factor ${\cal 
M}/\omega\cd p= {\cal M}/\omega\cd {\cal P}$ turns into $1/\omega_0$, 
and $\vec{\omega}/\omega_0$ gives $\vec{n}$. 

One can similarly represent the expressions containing the vector 
products. For example:
\begin{eqnarray}\label{eq85}
\vec{\sigma}_{12}\cd[\vec{q}_1\times\vec{q}_2]&\rightarrow&
-\epsilon^{\mu\nu\rho\gamma}
c_1c_2[\bar{u}(k_1)\Pi_+\gamma_{\mu}\Pi_- U_c\bar{u}(k_2)]
k_{1\nu}k_{2\rho}{\cal P}_{\gamma}/{\cal M},
\nonumber\\
\vec{\sigma}_{3N}\cd[\vec{q}_1\times\vec{q}_2]&\rightarrow&
-\epsilon^{\mu\nu\rho\gamma}
c_3c_N[\bar{u}(k_3)\Pi_+\gamma_{\mu}\gamma_5\Pi_+ u(p)]
k_{1\nu}k_{2\rho}{\cal P}_{\gamma}/{\cal M},
\nonumber\\
\vec{\sigma}_{12}\cd[\vec{q}_1\times\vec{n}]&\rightarrow&
-\epsilon^{\mu\nu\rho\gamma}
c_1c_2[\bar{u}(k_1)\Pi_+\gamma_{\mu}\Pi_-U_c\bar{u}(k_2)]
k_{1\nu}\omega_{\rho}{\cal P}_{\gamma}/\omega\cd p,
\nonumber\\
\vec{\sigma}_{3N}\cd[\vec{q}_1\times\vec{n}]&\rightarrow&
-\epsilon^{\mu\nu\rho\gamma}
c_3c_N[\bar{u}(k_3)\Pi_+\gamma_{\mu}\gamma_5\Pi_+ u(p)]
k_{1\nu}\omega_{\rho}{\cal P}_{\gamma}/\omega\cd p,
\end{eqnarray}
etc. By these substitutions the nucleon wave function can be easily 
represented in the four-dimensional form.  We do not give here the 
explicit formulas.

\section{Matrix elements of the current operator}\label{meq}
\subsection{The quark current}
The quark electromagnetic current has the form:
\begin{equation}\label{qc}
J_{\rho}=\sum_{i=1,2,3}
\bar{u}'_i\gamma_{\rho}u_i\;
\left[\xi^{\dag\tau'_i}_i\frac{1}{2}\left(\frac{1}{3}
+\tau_z^{(i)}\right)\xi^{\tau_i}_i \right]
\Rightarrow 
3\bar{u}'(k_3)\gamma_{\rho}u(k_3)\;\left[\xi^{\dag\tau'_3}_3\frac{1}{2} 
\left(\frac{1}{3}+\tau_z^{(3)}\right)
\xi^{\tau_3}_3\right],
\end{equation}
$\tau_z^{(i)}$ is the Pauli matrix acting on the spinor of the $i$-th 
quark.  The replacement by the only item has been made due to symmetry 
of the wave function to be used for calculating the matrix elements.  
In (\ref{qc}) the factor 
$$
\frac{1}{2}\left(\frac{1}{3}+\tau_z\right)=
\left(\begin{array}{cc} 2/3 & 0 \\ 0 & -1/3\end{array}\right)
$$
reproduces the charges of $u$ and $d$ quarks.

The quark axial current has the form:
\begin{equation}\label{ac}
J_{\rho}=\sum_{i=1,2,3}
\bar{u}'_i\gamma_{\rho}\gamma_5u_i\;
\left[\xi^{\dag\tau'_i}_i\vec{\tau}^{(i)}\xi^{\tau_i}_i\right]
\Rightarrow 3\bar{u}'(k_3)\gamma_{\rho}\gamma_5 u(k_3)\;
\left[\xi^{\dag\tau'_3}_3\vec{\tau}^{(3)}\xi^{\tau_3}_3\right],
\end{equation}

At first, we calculate the isospin matrix elements relative to the 
isospin wave functions $\chi^{(0)},\chi^{(1)}$ given by (\ref{eq24}).  
Let $I$ is a quark isotopic operator acting on the isospin indices of 
the quark 3. It is sandwiched with the isotopic spinors in (\ref{qc}) 
or in (\ref{ac}). Then we get:  
\begin{equation}\label{iso2} 
\langle 
{\chi}'^{(0)}|I|{\chi}^{(0)} \rangle 
=\sum_{\tau'_1\tau_1\tau_2\tau_3} [\xi^{\tau_2}_2\tau_y\xi^{\tau_1}_1] 
[\xi^{\dag\tau'}_N\xi^{\tau'_3}_3]
 I_{\tau'_3\tau_3}
[\xi^{\dag\tau_3}_3\xi^{\tau}_N]
[\xi^{\dag\tau_1}_1\tau_y\xi^{*\tau_2}_2]
=Tr[\tau_y\tau_y] I_{\tau'\tau} =2 I_{\tau'\tau},
\end{equation}
where $\tau,\tau'$ are the nucleon isospin projections.

For other matrix elements we similarly get:
\begin{equation}\label{iso4}
\langle {\chi}'^{(1)}|I|{\chi}^{(1)} \rangle  =
2(\tau_iI\tau_i)_{\tau'\tau},\quad
\langle {\chi}'^{(1)}|I|{\chi}^{(0)} \rangle 
=\langle {\chi}'^{(0)}|I|{\chi}^{(1)} \rangle=0.
\end{equation}
For the isospin structure of the electromagnetic current, eq. 
(\ref{qc}), we find:  
\begin{equation}\label{iso5}
\langle {\chi}'^{(0)}|I|{\chi}^{(0)} \rangle=
\left(\begin{array}{cc}
4/3 & 0 \\
0 & -2/3
\end{array}\right),\quad
\langle {\chi}'^{(1)}|I|{\chi}^{(1)} \rangle= \left(1
-\tau_z\right)_{\tau'\tau}=
\left(\begin{array}{cc}
0 & 0 \\
0 & 2
\end{array}\right).
\end{equation}
For the axial current $I=\vec{\tau}$ we obtain:
\begin{equation}\label{is6c}
\langle {\chi}'^{(0)}|I|{\chi}^{(0)} 
\rangle=2\vec{\tau}_{\tau'\tau}, \quad \langle 
 {\chi}'^{(1)}|I|{\chi}^{(1)} \rangle = 
2(\tau_i\vec{\tau}\tau_i)_{\tau'\tau}=
-2\vec{\tau}_{\tau'\tau}.
\end{equation}

\subsection{The spin matrix elements}
We calculate the spin matrix elements of the operator 
$J_{\rho}=\bar{u}'(k_3)j^{(3)}_{\rho}u(k_3)$ in the example of the 
symmetric wave function (\ref{eq86p}). For the electromagnetic current:  
$j^{(3)}_{\rho} =\gamma_{\rho}$.  For the axial current:  
$j^{(3)}_{\rho}=\gamma_{\rho}\gamma_5$.  

The final wave function reads (compare with (\ref{eq86p})):
\begin{eqnarray}\label{eq86a} 
{\mit\Psi}'^{\dagger}_S&=&\frac{\psi'^*_S}{\sqrt{72}} 
c'_1c'_2c'_3c'_N\{3[u(k_2)U_c\Pi'_-\gamma_5 \Pi'_+u(k_1)] 
[\bar{u}(p')\Pi'_+\Pi'_+ u(k'_3)] \nonumber\\ 
&&-[u(k_2)U_c\Pi'_-\gamma^{\mu}\Pi'_+ u(k_1)]
[\bar{u}(p')\Pi'_+\gamma_5\gamma_{\mu} \Pi'_+ u(k'_3)]
(\vec{\tau}_{21}\cd\vec{\tau}_{N3})\}.
\nonumber\\ 
&\equiv& {\mit \Psi}'^{\dagger}_0+{\mit \Psi}'^{\dagger}_1.
\end{eqnarray}
where
$$
\Pi'_{\pm}=\frac{{\cal M}'\pm\hat{\cal P}'}{2{\cal M'}},\quad
{\cal P}'=p'+\omega\tau'=k_1+k_2+k'_3,\quad 
{\cal M'}=\sqrt{{\cal P}'^2}
$$
and $c'_{1,2,3},c'_N$ differ from $c_{1,2,3},c_N$, eq. (\ref{eq81a}), 
by the replacement of the initial momenta by the final ones.  Since, 
due to (\ref{qc}) and (\ref{ac}), the quark 3 only is chosen as the 
interacting one, the momenta of the quarks 1 and 2 remain unchanged.  
The wave function ${\mit\Psi}'^{\dagger}_S$ is written for this case.

The matrix element can be represented as:
$$
\langle{\mit\Psi}'^{\dagger}_S|J_{\rho}|{\mit\Psi}_S\rangle =  
\langle{\mit\Psi}'^{\dagger}_0|J_{\rho}|{\mit\Psi}_0\rangle +
\langle{\mit\Psi}'^{\dagger}_1|J_{\rho}|{\mit\Psi}_1\rangle.
$$
The nondiagonal matrix elements between the states 
$\langle{\mit\Psi}'^{\dagger}_0|$ and $|{\mit\Psi}_1\rangle$ do 
not contribute because of the zero nondiagonal isospin matrix elements, 
eq.  (\ref{iso4}).

For the matrix element $\langle 
{\mit\Psi}'_0|J_{\rho}|{\mit\Psi}_0\rangle$ 
we start with the expression 
in terms of the spinors (omitting for a moment any isospin factors):  
\begin{eqnarray}\label{me0} 
\langle {\mit\Psi}'_0|J_{\rho}|{\mit\Psi}_0 \rangle&=& 
\frac{9}{72}
c\psi'^*_S\psi_S\sum_{\sigma_1,\sigma_2,\sigma_3,\sigma'_3} 
[u^{\sigma_2}(k_2)U_c\gamma_5\Pi'_+u^{\sigma_1}(k_1)]\;
[\bar{u}^{\sigma_1}(k_1)\Pi_+\gamma_5 U_c\bar{u}^{\sigma_2}(k_2)]
\nonumber\\
&\times&
[\bar{u}^{\sigma'}(p')\Pi'_+ u^{\sigma'_3}(k'_3)]
[\bar{u}^{\sigma'_3}(k'_3) j^{(3)}_{\rho} u^{\sigma_3}(k_3)]\;
[\bar{u}^{\sigma_3}(k_3)\Pi_+u^{\sigma}(p)]/x_3,
\end{eqnarray}
where 
$$
c=c_1 c_2 c_3 c_N c'_1 c'_2 c'_3 c'_N,\quad
x_3=\omega\cd k_3/\omega\cd p. 
$$
We take into account that 
$\sum_{\sigma}u^{\sigma}_{\alpha'}(k)\bar{u}^{\sigma}_{\alpha}(k)= 
(\hat{k}+m)_{\alpha'\alpha}$, $(U_c \gamma_5)^t= - \gamma_5 U_c= 
-U_c\gamma_5$ and $\gamma_5 U_c(\hat{k}+m)^t=(\hat{k}+m)\gamma_5 U_c$ 
and get:
\begin{eqnarray}\label{sp1p}
\langle {\mit\Psi}'_0|J_{\rho}|{\mit\Psi}_0 \rangle&=& 
\frac{9}{72}c\psi'^*_S\psi_S Tr[\Pi'_+(\hat{k}_1+m)\Pi_+(\hat{k}_2+m)]
\nonumber\\
&\times&
[\bar{u}^{\sigma'}(p')\Pi'_+ (\hat{k}'_3+m)j^{(3)}_{\rho}(\hat{k}_3+m) 
\Pi_+u^{\sigma}(p)]/x_3.
\end{eqnarray}
Similarly we get another matrix element:
\begin{eqnarray}\label{me1}
\langle {\mit\Psi}'_1|J_{\rho}|{\mit\Psi}_1 \rangle&=& 
\frac{1}{72}c\psi'^*_S\psi_S Tr[\Pi'_+\gamma_5
\gamma^{\nu}\Pi'_+(\hat{k}_1+m)\Pi_+\gamma^{\mu}\gamma_5\Pi_+
(\hat{k}_2+m)]
\nonumber\\
&\times&
\bar{u}^{\sigma'}(p') 
\Pi'_+\gamma_5\gamma_{\nu}\Pi'_+(\hat{k}'_3+m)
j^{(3)}_{\rho}(\hat{k}_3+m)
\Pi_+\gamma_{\mu} \gamma_5\Pi_+u^{\sigma}(p)]/x_3.
\end{eqnarray}
Taking into account the isospin factors (\ref{iso5}),  we find by this 
way the matrix elements of the nucleon electromagnetic current.

{\bf For proton:}
\begin{equation}\label{me3}
\langle p'|J_{\rho}|p\rangle =3 \int  \{\frac{4}{3}
\langle {\mit\Psi}'_0|J_{\rho}|{\mit\Psi}_0\rangle\} D.
\end{equation}
Note that the matrix element 
$\langle{\mit\Psi}'_1|J_{\rho}|{\mit\Psi}_1\rangle$ 
does not contribute here.

{\bf For neutron:}
\begin{equation}\label{me4}
\langle n'|J_{\rho}|n\rangle = 3\int  \{
-\frac{2}{3}\langle {\mit\Psi}'_0|J_{\rho}|{\mit\Psi}_0\rangle 
+2\langle {\mit\Psi}'_1|J_{\rho}|{\mit\Psi}_1\rangle \}D.
\end{equation}

With the isospin factor (\ref{is6c}) we get the matrix elements 

{\bf for the axial current:}
\begin{eqnarray}\label{me5}
\langle N'|J^A_{\rho}|N\rangle &=& \vec{\tau}A_{\rho},
\nonumber\\
A_{\rho}&=& 3\int  \{
2\langle {\mit\Psi}'_0|J_{\rho}|{\mit\Psi}_0\rangle 
-2\langle {\mit\Psi}'_1|J_{\rho}|{\mit\Psi}_1\rangle \}D.
\end{eqnarray}
These expressions can be represented in the form:
\begin{equation}\label{me4a}
\langle N'|J_{\rho}|N\rangle =\bar{u}'(p')G_{\rho}u(p)
\end{equation}
and determine by this way the $4\times 4$-matrix $G_{\rho}$.
Then we find the matrix 
\begin{equation}\label{ax2}
O_{\rho}=(\hat{p}'+M)G_{\rho}(\hat{p}+M)/(4M^2).
\end{equation}
The electromagnetic form factors are expressed through $O_{\rho} $
by the formulas given in the paper \cite{km96}. 

\section{Some applications}\label{appl}
\subsection{The proton magnetic moment}\label{ge}
The nucleon electromagnetic vertex in the light-front dynamics, i.e., 
the matrix $G_{\rho}$ in (\ref{me4a}), due to approximations, depends 
on the orientation of the light-front plane. In the covariant LFD this 
dependence is parametrized in terms of the four-vector $\omega$.  The 
vertex contains the non-physical $\omega$-depending terms, namely 
\cite{km96}:
\begin{eqnarray}\label{eq12}
G_{\rho}&=&F_1\gamma_{\rho}
     +\frac{iF_2}{2M}\sigma_{\rho\nu}q^{\nu}
+B_1\left(\frac{\hat{\omega}}{\omega\cd p} -\frac{1}{(1+\eta)M}
\right)P_{\rho}
+B_2
\frac{M}{\omega\cd p}\omega_{\rho}
+B_3
\frac{M^2}{(\omega\cd p)^2}\hat{\omega}\omega_{\rho}\ .
\nonumber\\
& &
\end{eqnarray}
The method of calculating $G_{\rho}$ is explained in the previous 
section. The physical form factors $F_1,F_2$ should be extracted from 
$G_{\rho}$ so to separate them from the nonphysical terms $B_{1-3}$. 

Usually the form factors are found from the plus-component of the 
current, corresponding to the contraction $G_{\rho}\omega^{\rho}$.
In this way, 
the terms $B_{1-3}$  do not contribute to the charge form factor 
$G_E=F_1-\eta F_2$ (here $\eta=Q^2/(4M^2)$).  It is given by the 
formula \cite{km96}:
\begin{equation}\label{me4b} 
G_E=\frac{M}{2(\omega\cd p)}Tr[O_{\rho}\omega^{\rho}].  
\end{equation} 
With the matrix $O_{\rho}$ determined through (\ref{me3}), (\ref{me4a}) 
and (\ref{ax2}) one can easily check that the proton form factor 
$G_E(0)$ coincides with the normalization integral (\ref{eq86b}).  With 
$O_{\rho}$ determined from (\ref{me4}) the neutron form factor $G_E(0)$ 
equals to zero.

However, the form factor $F_2$ calculated through the plus-component 
(we denote it as $F_2'$) contains the contribution of $B_1$, namely:
\begin{equation}\label{f2}
F_2'=F_2+\frac{2}{1+\eta}B_1=
 \frac{M}{2\eta(\omega\cd p)}
Tr\left[O_{\rho}\omega^{\rho}
\left(\hat{\omega}\frac{M}{\omega\cd p} -1\right)\right] .  
\end{equation} 
The formula for the form factor $F_2$, 
separated from $B_1$, is more lengthy. It is given in \cite{km96}.

The proton magnetic moment is $\mu_p=G_M(0)$ and the anomalous magnetic 
moment is $a=G_M(0)-1= F_2(0)$.  We analyze two cases: ({\it i}) 
the form factors calculated through the plus-component of the current, 
without separating the non-physical $\omega$-depending terms in the 
nucleon electromagnetic vertex (that in the covariant approach 
corresponds to the contraction $G_{\rho}\omega^{\rho}$); ({\it ii}) 
with separating the $\omega$-depending terms.  In the first case we 
find from (\ref{f2}) for the anomalous magnetic moment:  
\begin{equation}\label{me10} 
a'=\frac{2M}{{\cal M}}\int\left[\frac{(1-x_3){\cal M}(m+x_3{\cal M}) 
-\vec{R}^2_{\perp 3}/2} {(m+x_3 {\cal M})^2+\vec{R}^2_{\perp 
3}}\right]\psi_S^2 D.  
\end{equation} 
The expression for $a'$ is written in the variables explained above in 
the section \ref{norm1}.  It coincides with  one found in \cite{cc91} 
(see also \cite{brs94}).  Note that in these papers it is result of 
taking into account the Melosh rotation matrices.  In the covariant 
approach the spin rotations are included automatically.

In the case ({\it ii}), i.e., separating the $\omega$-dependent 
terms, we get:
\begin{equation}\label{me11}
a=2\int\left[\frac{(1-x_3)(m+x_3{\cal 
M})^2 -2x_3\vec{R}^2_{\perp 3}} {2x_3[(m+x_3 {\cal 
M})^2+\vec{R}^2_{\perp 3}]}\right]\psi_S^2 D 
\end{equation}
It differs from (\ref{me10}) by the relativistic corrections which can 
be important. It does not contain explicitly the nucleon mass.  In the 
nonrelativistic limit $M={\cal M}=3m$, $x_3=1/3$, $\vec{R}_{\perp 3}=0$ 
we find:  $a'=a=2$.

Note that the above expressions, for example, eq. (\ref{me4b}), 
contain the product of two traces, since according to (\ref{sp1p}) 
$O_{\rho}$ includes already the trace over the loop of the quarks 1 and 
2. This factorization is due to the fact that after permutations we 
have reduced the wave function to the form with the contracted pairs of 
the spinors 1-2 and 3-$N$.  Otherwise, instead of the product of traces 
we would get trace of the product of the larger number of the Dirac 
matrices.  This would complicate the calculations considerably.

Calculating, for example, eq. (\ref{me10}) through the trace 
(\ref{f2}), we get the expression in terms of the scalar products 
between the four-vectors entering in (\ref{f2}) (initial and final 
nucleon and quark momenta $p,p',k_1,k_2,k_3,k'_3$ and also 
$\omega,{\cal P},{\cal P}'$). Then we express these scalar products in 
terms of the variables $\vec{R}_{\perp i}, \vec{R}_{\perp i}', x_i$. 
These three-body kinematical relations are given in appendix to the 
paper \cite{km96}.\footnote{In the paper \cite{km96} the quark 1 was 
considered as the interacting one, whereas in the present paper it is 
the quark 3.  Therefore in the latter case instead of the formulas of 
\cite{km96}  
$$
\vec{R}_{\perp 1}'=\vec{R}_{\perp 1}+(1-x_1)\vec{\Delta},\quad
\vec{R}_{\perp 2,3}'=\vec{R}_{\perp 2,3}-x_{2,3}\vec{\Delta}
$$
one should use
$$
\vec{R}_{\perp 1,2}'=\vec{R}_{\perp 1,2}-x_{1,2}\vec{\Delta},\quad
\vec{R}_{\perp 3}'=\vec{R}_{\perp 3}+(1-x_3)\vec{\Delta}.
$$}

\subsection{The axial form factor}\label{axial}
The vertex function corresponding to the axial current in the impulse 
approximation is given by eq. (\ref{me5}). As in the case of the 
nucleon electromagnetic vertex, it also depends on the orientation of 
the light-front plane.  In the particular case of $q=0$ it has the 
form:
\begin{eqnarray}\label{ax1}
A_{\rho}&=&\bar{u}'(p)G^A_{\rho}u(p),
\nonumber\\
G^A_{\rho}& =&\left(g_A(0)\gamma_{\rho}+B_{14}(0)\frac{1}{\omega\cd p}
\hat{\omega}p_{\rho}+B_3(0)\frac{M^2}{(\omega\cd p)^2}\hat{\omega}
\omega_{\rho}\right)\gamma_5
\end{eqnarray}
From eq. (\ref{ax1}) we find:
\begin{equation}\label{me7}
g_A(0)=Tr\{O_{\rho}[\hat{\omega}p^{\rho}/ (\omega\cd p) -
\hat{\omega}\omega^{\rho}M^2/(\omega\cd p)^2-
\gamma^{\rho}]\gamma_5\}/4,
\end{equation}
where $O_{\rho}$ is given by (\ref{ax2}) at $p'=p$ with the replacement 
$G_{\rho} \rightarrow G^A_{\rho}$. 
Since the expression (\ref{me7}), besides $\omega_{\rho}$, contains the 
contractions with $p^{\rho}$ and $\gamma^{\rho}$, this means that
like in the case of $F_2$ the plus-component is not enough to find 
$g_A$.

However, following the standard procedure and contracting $A_{\rho}$ 
with $\omega_{\rho}$, we get:
\begin{equation}\label{me6}
g'_A(0)=g_A(0)+B_{14}(0)=\frac{M^2}{2(\omega\cd p)^2}
Tr[O_{\rho}\omega^{\rho}\hat{\omega}\gamma_5].
\end{equation}
Finding $G_{\rho}^A$ from (\ref{me5}) and substituting it in 
(\ref{ax2}) we get $O_{\rho}$. Then by (\ref{me6}) we find:  
\begin{equation}\label{me8}
g'_A(0)= \frac{5}{3}\int\left[\frac{(m+x_3 {\cal M})^2-\vec{R}^2_{\perp 
3}} {(m+x_3 {\cal M})^2+\vec{R}^2_{\perp 3}}\right]\psi_S^2 D.
\end{equation}
This expression coincides with one found the papers \cite{cc91,ma} (see 
also \cite{brs94}). Like the magnetic moment, in these papers it also 
results from  the Melosh rotation matrices. 

Substituting $O_{\rho}$ in eq. (\ref{me7}) we get $g_A(0)$ separated 
from the nonphysical form factor $B_{14}(0)$:  
\begin{equation}\label{me9}
g_A(0)= \frac{5}{3}\int\left[ \frac{(m+x_3 {\cal M})
(m^2+x_3m {\cal M}+\vec{R}^2_{\perp 3})} 
{x_3 M[(m+x_3 {\cal M})^2+\vec{R}^2_{\perp 3}]}\right]\psi_S^2 D.
\end{equation}
In the nonrelativistic limit we find: $g'_A(0)=g_A(0)=5/3$. 

\section{Conclusion}\label{concl}
We have established the general spin structure of the nucleon wave 
function in $3q$-model. It contains sixteen spin components forming the 
full basis for decomposition of the wave function.  After taking into 
account the isospin, the total number of all the scalar functions 
$\psi_S,\psi_A, \psi_{1,2}$, which are the coefficients at the front of 
all the structures, namely, the momentum-independent ones, sect.  
\ref{indep}, the structures generated by the symmetric tensors, 
appendix \ref{appwf1}, and, at last, the structures generated by the 
antisymmetric tensors, appendix \ref{appwf2}, is 32.  This just 
corresponds to 16 functions for proton and 16 for neutron.

We give recipe representing all these structures in the explicitly 
covariant form.  After that the calculations of the form factors are 
the straightforward, through the standard routine of the Dirac matrices 
and of the trace techniques, like in the case of the Feynman approach.  
This fact is the strong advantage of the explicitly covariant version 
of LFD. 

Among four spin structures independent of momenta there is one 
symmetric structure, eq. (\ref{eq27}). Namely it, multiplied by 
the Melosh matrices, with the function $\psi_S$ approximated by 
the S-wave, is often used for calculating the nucleon properties (form 
factors, magnetic moments, etc.).  Represented in the covariant form by 
eq. (\ref{eq86p}), it does not require any Melosh matrices.  Our 
calculations by this way of the proton anomalous magnetic moment and 
of the nucleon axial form factor, with the same wave function eq.  
(\ref{eq27}), like in the paper \cite{cc91}, but represented  
covariantly, gives, of course, the same result as in \cite{cc91}.

Separating the $\omega$-dependent terms in the vector and axial 
vertices, but still keeping only one spin structure, we get different 
results. We believe that in order to make comparison with experiment, 
one should investigate the influence of other components, including the 
new ones.  Some of them may be important.  One can either introduce 
these components phenomenologically and see their value from fit of the 
experimental data, or try to estimate them in an appropriate quark 
dynamics.  Even in the deuteron wave function we found \cite{ck-deut} 
the component (so called $f_5$) which dominates in moderately 
relativistic region over all other components, including S- and 
D-waves. The same can take place in the nucleon wave function.

\section*{Acknowledgement}
The author is sincerely grateful to M. Beyer, V.M.  Kolybasov, J.-F. 
Mathiot and H.J. Weber for stimulating discussions and valuable 
remarks.

\appendix
\section{The Dirac matrices and the spinors}\label{appen1}
We give below the explicit form of the Dirac matrices and the spinors 
in the standard representation.

The nucleon spinor has the form:
\begin{equation}\label{eq74} 
u^{\sigma}(p)= \sqrt{\varepsilon_{p}+M} \left(\begin{array}{c} 
1\\
\frac{\displaystyle\vec{\sigma}\cd\vec{p}}
{\displaystyle(\varepsilon_p+M)}
\end{array}\right) w_N^{\sigma}, 
\end{equation}
where $w_N^{\sigma}$ is the two-component spinor and $\varepsilon_p 
= \sqrt{\vec{p}\,^2+M^2}$.  

The spinor $\bar{u}(k_1)$ of the quark 1 reads:  
\begin{equation}\label{eq75}
\bar{u}^{\sigma_1} (k_1)=u^{\dagger\sigma_1}(k)\gamma^0
=\sqrt{\varepsilon_{k_1}+m}\;w_1^{\dagger\sigma_1}
\left(1,\; -\frac{\displaystyle\vec{\sigma}\cd\vec{k}_1}
{\displaystyle(\varepsilon_{k_1}+m)}\right)
\end{equation}                                 
and similarly for  $\bar{u}(k_3)$.

For the quark 2 we use the charge conjugated spinor:
\begin{equation}\label{eq76} 
U_c\bar{u}^t(k_2)=
\sqrt{\varepsilon_{k_2}+m} 
\left(\begin{array}{c} 
\frac{\displaystyle{-\vec{\sigma}\cd\vec{k}_2}}
{\displaystyle(\varepsilon_{k_2}+m)}
\\
-1
\end{array}\right) \sigma_y w_2^*. 
\end{equation}
The charge conjugation matrix $U_c=\gamma^2 \gamma^0$:
$$U_c=
\left(
\begin{array}{ll}
\quad 0 & -\sigma_y\\
-\sigma_y &\quad 0
\end{array}
\right)
$$
The Dirac matrices:
\begin{equation}\label{dg}
\gamma^0 = \left(\begin{array}{rr} 1 & 0 \\ 0 & -1 
\end{array}\right),\quad \vec{\gamma}= 
\left(\begin{array}{rr}0&\vec{\sigma} \\ -\vec{\sigma} & 0 
\end{array}\right),\quad
\gamma_5 =-i\gamma^0\gamma^1\gamma^2\gamma^3 =
\left(\begin{array}{rr}0&-1\\-1&0\end{array}\right),
\end{equation}                                                                  
where $\vec{\sigma}$ are the Pauli matrices. 

\section{Relation between the tensors}\label{appen2}
The relation between the symmetric traceless tensors of the second rank,
defined by (\ref{eq46}), has the form:  
\begin{eqnarray}\label{eq31} 
t^{ij}\equiv &&\quad 
(1-\cos^2\theta_{2n})T_{11}^{ij}/\vec{q}\,_1^2+
(1-\cos^2\theta_{1n})T_{22}^{ij}/\vec{q}\,_2^2+
(1-\cos^2\theta_{12})T_{nn}^{ij}
\nonumber\\
&& - (\cos\theta_{12}
-\cos\theta_{1n}\cos\theta_{2n})T_{12}^{ij}/(|\vec{q}_1||\vec{q}_2|)
\nonumber\\
&&-(\cos\theta_{1n}
-\cos\theta_{12}\cos\theta_{2n})T_{1n}^{ij}/|\vec{q}_1|
\nonumber\\
&&-(\cos\theta_{2n}
-\cos\theta_{12}\cos\theta_{1n})T_{2n}^{ij}/|\vec{q}_2|
\equiv 0.
\end{eqnarray}
where
$$\cos\theta_{12}=\vec{q}_1\cd\vec{q}_2/(|\vec{q}_1||\vec{q}_2|),\quad
\cos\theta_{1n}=\vec{q}_1\cd\vec{n}/|\vec{q}_1|,\quad
\cos\theta_{2n}=\vec{q}_2\cd\vec{n}/|\vec{q}_2|.
$$
In order to prove (\ref{eq31}), it is enough to check that:
\begin{eqnarray}\label{eq32}
&& q_1^i t_{ij}q_1^j=0,\quad q_2^i t_{ij}q_2^j=0,\quad 
n^i t_{ij}n^j=0, 
\nonumber\\ 
&& q_1^i t_{ij}q_2^j=0,\quad 
q_1^i t_{ij}n^j=0,\quad q_2^i t_{ij} n^j=0.  
\end{eqnarray} 
Since any vectors $\vec{a}, \vec{b}$ can be decomposed in terms of 
$\vec{q}_1,\vec{q}_2,\vec{n}$, e.g.:  $$ 
\vec{a}=c_1\vec{q}_1+c_2\vec{q}_2+c_3\vec{n},
$$
from (\ref{eq32}) it follows that $a^i t_{ij}b^j\equiv 0$ for arbitrary
$\vec{a}, \vec{b}$.
This means that 
$$
t_{ij}\equiv 0.
$$
The identity (\ref{eq31}) allows to choose as a basis any five tensors.
We exclude the tensor $T_{nn}^{ij}$.

\section{Relation between the spin structures}\label{appen3}
It is not difficult to establish the following identities:
\begin{equation}\label{ap1}
\begin{array}{llll}
(\vec{\sigma}\cd\vec{q}_1)\;(\vec{n}\cd [\vec{q}_1\times\vec{q}_2])
&\!\!=\vec{\sigma}\cd[\vec{q}_1\times\vec{q}_2]\; 
(\vec{q}_1\cd\vec{n})
&\!\!-\vec{\sigma}\cd[\vec{q}_1\times\vec{n}]\; 
(\vec{q}_1\cd\vec{q}_2)
&\!\!+\vec{\sigma}\cd[\vec{q}_2\times\vec{n}]\; \vec{q}\,_1^2
\\ 
(\vec{\sigma}\cd\vec{q}_2)\;(\vec{n}\cd [\vec{q}_1\times\vec{q}_2])
&\!\!=\vec{\sigma}\cd[\vec{q}_1\times\vec{q}_2]\; (\vec{q}_2\cd\vec{n})
&\!\!-\vec{\sigma}\cd[\vec{q}_1\times\vec{n}]\; \vec{q}\,_2^2
&\!\!+\vec{\sigma}\cd[\vec{q}_2\times\vec{n}]\; (\vec{q}_1\cd\vec{q}_2)
\\
(\vec{\sigma}\cd\vec{q}_3)\;(\vec{n}\cd [\vec{q}_1\times\vec{q}_2])
&\!\!=\vec{\sigma}\cd[\vec{q}_1\times\vec{q}_2] 
&\!\!-\vec{\sigma}\cd[\vec{q}_1\times\vec{n}]\; (\vec{q}_2\cd\vec{n})
&\!\!+\vec{\sigma}\cd[\vec{q}_2\times\vec{n}]\; (\vec{q}_1\cd\vec{n})
\end{array}
\end{equation}
where $\vec{\sigma}$ is arbitrary vector.

The reverse expressions have the form:
\begin{eqnarray}\label{app1}
\vec{\sigma}\cd[\vec{q}_1\times\vec{q}_2]&=& 
(\vec{n}\cd [\vec{q}_1\times\vec{q}_2])\;
\{-(\vec{\sigma} \cd\vec{q}_1) [\vec{q}\,_2^2 (\vec{q}_1\cd\vec{n})
-(\vec{q}_1\cd\vec{q}_2)(\vec{q}_2\cd\vec{n})]
\nonumber\\
&-&(\vec{\sigma} \cd\vec{q}_2) 
[\vec{q}\,_1^2 (\vec{q}_2\cd\vec{n}) - 
(\vec{q}_1\cd\vec{q}_2)(\vec{q}_1\cd\vec{n})]
+(\vec{\sigma} \cd\vec{n}) 
[\vec{q}\,_1^2 \vec{q}\,_2^2  - 
(\vec{q}_1\cd\vec{q}_2)^2 ]\}/d
\nonumber\\
\vec{\sigma}\cd[\vec{q}_1\times\vec{n}]&=& 
(\vec{n}\cd [\vec{q}_1\times\vec{q}_2])\;
\{(\vec{\sigma} \cd\vec{q}_1) [\vec{q}_1\cd\vec{q}_2  
-(\vec{q}_1\cd\vec{n})(\vec{q}_2\cd\vec{n})] 
\nonumber\\
&-&(\vec{\sigma} \cd\vec{q}_2) 
[\vec{q}\,_1^2 - (\vec{q}_1\cd\vec{n})^2]
+(\vec{\sigma} \cd\vec{n}) 
[\vec{q}\,_1^2 (\vec{q}_2\cd\vec{n}) - 
(\vec{q}_1\cd\vec{q}_2) (\vec{q}_1\cd\vec{n})]\}/d
\nonumber\\
\vec{\sigma}\cd[\vec{q}_2\times\vec{n}]&=& 
(\vec{n}\cd [\vec{q}_1\times\vec{q}_2])\;
\{(\vec{\sigma} \cd\vec{q}_1) [\vec{q}\,_2^2  
-(\vec{q}_2\cd\vec{n})^2] 
\nonumber\\
&-&(\vec{\sigma} \cd\vec{q}_2) 
[\vec{q}_1\cd\vec{q}_2 - (\vec{q}_1\cd\vec{n})(\vec{q}_2\cd\vec{n})]
-(\vec{\sigma} \cd\vec{n}) 
[\vec{q}\,_2^2 (\vec{q}_1\cd\vec{n}) - 
(\vec{q}_1\cd\vec{q}_2) (\vec{q}_2\cd\vec{n})]\}/d
\end{eqnarray}
where $$
d=(\vec{n}\cd [\vec{q}_1\times\vec{q}_2]) ^2=
\vec{q}\,_1^2 \vec{q}\,_2^2 - \vec{q}\,_1^2 (\vec{q}_2\cd\vec{n})^2
- \vec{q}\,_2^2 (\vec{q}_1\cd\vec{n})^2
-(\vec{q}_1\cd\vec{q}_2)^2 +
2 (\vec{q}_1\cd\vec{q}_2) (\vec{q}_1\cd\vec{n}) (\vec{q}_2\cd\vec{n}) 
$$
Due to these identities the set of the structures ({\it 11-16}) of 
(\ref{eq34c}) is equivalent to the set formed by the structures 
$(\vec{\sigma}_{12}\cd\vec{q}_1)(\vec{n}\cd[\vec{q}_1\times\vec{q}_2])$, 
$(\vec{\sigma}_{12}\cd\vec{q}_2)(\vec{n}\cd[\vec{q}_1\times\vec{q}_2])$,
etc.

\section{The Fierz identities for the Pauli matrices}\label{Fierz}
To establish the permutation properties (\ref{eq20}) and (\ref{eq53b}) 
of the spin operators relative to the permutation $P_{13}$, we use the 
following Fierz identities:  
\begin{eqnarray}
P_{13}1_{12}1_{3N}=
1_{32}1_{1N}&=&\frac{1}{2}1_{12}1_{3N}+
\frac{1}{2}\vec{\sigma}_{12}\cd\vec{\sigma}_{3N},
\nonumber\\
P_{13}1_{12}\sigma^i_{3N}=
1_{32}\sigma^i_{1N}&=&\frac{1}{2}1_{12}\sigma^i_{3N}
+\frac{1}{2}\sigma^i_{12}1_{3N}
-\frac{i}{2}\epsilon_{ikl}\sigma^k_{12}\sigma^l_{3N},
\nonumber\\
P_{13}\sigma^i_{12}1_{3N}=
\sigma^i_{32}1_{1N}&=&\frac{1}{2}1_{12}\sigma^i_{3N}
+\frac{1}{2}\sigma^i_{12}1_{3N}
+\frac{i}{2}\epsilon_{ikl}\sigma^k_{12}\sigma^l_{3N},
\nonumber\\
P_{13}\sigma^i_{12}\sigma^j_{3N}=
\sigma^i_{32}\sigma^j_{1N}&=&
\frac{1}{2}\delta_{ij}1_{12}1_{3N}
+\frac{i}{2}\epsilon_{ijl}1_{12}\sigma^l_{3N}
-\frac{i}{2}\epsilon_{ijl}\sigma^l_{12}1_{3N}
\nonumber\\
&-&\frac{1}{2}\delta_{ij}\vec{\sigma}_{12}\cd\vec{\sigma}_{3N}
+\frac{1}{2}\sigma^i_{12}\sigma^j_{3N}
+\frac{1}{2}\sigma^j_{12}\sigma^i_{3N}.
\label{eq41}
\end{eqnarray} 
From (\ref{eq41}) it follows:
\begin{equation}
P_{13}\vec{\sigma}_{12}\cd\vec{\sigma}_{3N}=
\vec{\sigma}_{32}\cd\vec{\sigma}_{1N}=
\frac{3}{2} 1_{12}1_{3N}
-\frac{1}{2}\vec{\sigma}_{12}\cd\vec{\sigma}_{3N}.
\nonumber\\
\end{equation}
Under the permutation $P_{23}$ the operators are transformed as 
follows:
\begin{eqnarray}
P_{23}1_{12}1_{3N}=
1_{13}1_{2N}&=&\quad\!\frac{1}{2}1_{12}1_{3N}
-\frac{1}{2}\vec{\sigma}_{12}\cd\vec{\sigma}_{3N},
\nonumber\\
P_{23}1_{12}\sigma^i_{3N}=
1_{13}\sigma^i_{2N}&=&\quad\!
\frac{1}{2}1_{12}\sigma^i_{3N}
-\frac{1}{2}\sigma^i_{12}1_{3N}
+\frac{i}{2}\epsilon_{ikl}\sigma^k_{12}\sigma^l_{3N},
\nonumber\\
P_{23}\sigma^i_{12}1_{3N}=
\sigma^i_{13}1_{2N}&=&
-\frac{1}{2}1_{12}\sigma^i_{3N}
+\frac{1}{2}\sigma^i_{12}1_{3N}
+\frac{i}{2}\epsilon_{ikl}\sigma^k_{12}\sigma^l_{3N},
\nonumber\\
P_{23}\sigma^i_{12}\sigma^j_{3N}=
\sigma^i_{13}\sigma^j_{2N}&=&
-\frac{1}{2}\delta_{ij}1_{12}1_{3N}
-\frac{i}{2}\epsilon_{ijl}1_{12}\sigma^l_{3N}.
-\frac{i}{2}\epsilon_{ijl}\sigma^l_{12}1_{3N}
\nonumber\\
&&-\frac{1}{2}\delta_{ij}\vec{\sigma}_{12}\cd\vec{\sigma}_{3N}
+\frac{1}{2}\sigma^i_{12}\sigma^j_{3N}
+\frac{1}{2}\sigma^j_{12}\sigma^i_{3N}.
\label{eq41a}
\end{eqnarray} 
From (\ref{eq41a}) it follows:
\begin{equation}
P_{23}\vec{\sigma}_{12}\cd\vec{\sigma}_{3N}=
\vec{\sigma}_{13}\cd\vec{\sigma}_{2N}=
-\frac{3}{2} 1_{12}1_{3N}
-\frac{1}{2}\vec{\sigma}_{12}\cd\vec{\sigma}_{3N}.
\nonumber\\
\end{equation}
The permutation $P_{12}$ is evident:
\begin{equation}\label{eq43a} 
P_{12}1_{12}= 1_{21}=-1_{12},\quad 
P_{12}\vec{\sigma}_{12}=\vec{\sigma}_{21}=\vec{\sigma}_{12}.
\end{equation}
A set of the Fierz identities for the Dirac matrices is given in 
\cite{weber}.

\section{Wave functions generated by the symmetric tensor structures}
\label{appwf1}
Below we give the structures generated by the symmetric tensors, as 
described in sect. \ref{symm}. 
The diagram shown below summarizes all the ways of generating the 
nucleon wave function ${\mit \Psi}_S$ from the functions of different 
symmetries.

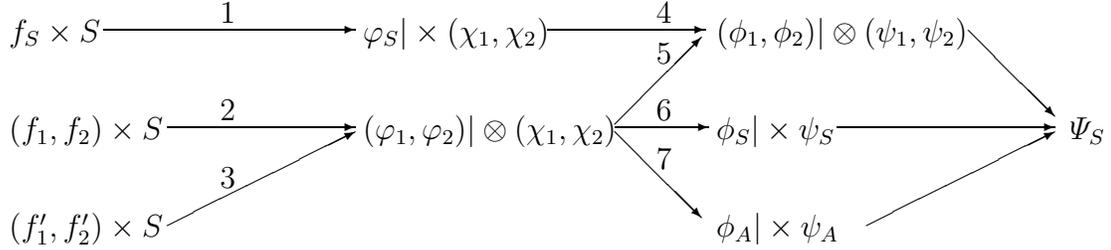
\begin{picture}(150,41)
\put(2,31){$f_S\times S$}
\put(2,18){$(f_1,f_2)\times S$}
\put(2,5){$(f'_1,f'_2)\times S$}

\put(49,31){$\varphi_S|\times (\chi_1,\chi_2)$}
\put(49,18){$(\varphi_1,\varphi_2)|\otimes (\chi_1,\chi_2)$}

\put(96,31){$(\phi_1,\phi_2)|\otimes (\psi_1,\psi_2)$}
\put(96,18){$\phi_S|\times\psi_S$}
\put(96,5){$\phi_A|\times\psi_A$}

\put(143,18){${\mit \Psi}_S$}

\put(14.5,32.5){\vector(1,0){33.5}}
\put(23,19.5){\vector(1,0){25}}
\put(23,6.5){\vector(2,1){25}}

\put(30,33.5){1}
\put(30,20.5){2}
\put(30,11.5){3}

\put(73.5,32.5){\vector(1,0){21}}
\put(82.5,20){\vector(1,1){11.5}}
\put(82.5,19.5){\vector(1,0){12.5}}
\put(82.5,19){\vector(1,-1){11.5}}

\put(88,33.5){4}
\put(88,28){5}
\put(88,20.5){6}
\put(88,14){7}

\put(129.5,32.5){\vector(1,-1){11.5}}
\put(112,19.5){\vector(1,0){29}}
\put(116,6.5){\vector(2,1){25}}
\end{picture}

For example, the path $2\rightarrow 5$ means that from the doublet 
$(f_1,f_2)$ multiplied by the symmetric spin tensor $ S$ we obtain 
the doublet $(\varphi_1,\varphi_2)$. Then with the isospin doublet 
$(\chi_1,\chi_2)$, by eq. (\ref{eq14c},c), we construct the 
spin-isospin doublet $(\phi_1,\phi_2)$. Then with the momentum scalar 
functions $(\psi_1,\psi_2)$, by (\ref{eq9}), we get ${\mit \Psi}_S$ 
given by (\ref{eq53p}). Seven paths connecting the initial states at 
the left of the diagram with the final one ${\mit \Psi}_S$ correspond 
to seven functions given below.

Before any function we indicate in more detail, how the given structure 
is obtained. For example, the notation above eq. (\ref{eq50p})
$$
2\rightarrow 6:\qquad
\left.\left.
\stackrel{(\ref{eq48},a)}{(f_1,f_2)}\times \stackrel{(\ref{eq44})}
{ S}
\stackrel{(\ref{eq44a})}{\longrightarrow}
(\varphi_1,\varphi_2)\right|\otimes 
\stackrel{(\ref{eq23})}{(\chi_1,\chi_2)}
\stackrel{(\ref{eq9cc})}{\longrightarrow}
\phi_S\right|\times \psi_S
\stackrel{(\ref{eq8})}{\longrightarrow}
\stackrel{(\ref{eq50p})}{{\mit \Psi}_S}
$$
means that eq. (\ref{eq50p}) is obtained by the path $2\rightarrow 6$.  
Namely, the doublet (\ref{eq48},a), formed by the momentum depending 
functions $(f_1,f_2)$, is multiplied by the symmetric spin tensor ${ 
S}$, defined by (\ref{eq44}), and gives by (\ref{eq44a}) the spin 
doublet $(\varphi_1,\varphi_2)$ (the line 2 on the diagram).  Then, 
together with the isospin doublet $(\chi_1,\chi_2)$, defined by 
(\ref{eq23}), by means of eq. (\ref{eq9cc}), it gives the symmetric 
spin-isospin function $\phi_S$ (the line 6 on the diagram).  Multiplied 
by $\psi_S$ by eq. (\ref{eq8}), it gives the total symmetric nucleon 
wave function ${\mit \Psi}_S$, eq. (\ref{eq50p}). All other notations 
are similar.  
\begin{eqnarray} 
\hline
\nonumber\\
1\rightarrow 4:
&&
\left.\left.
\stackrel{(\ref{eq47a})}{f_S}\times \stackrel{(\ref{eq44})}
{ S}
\stackrel{(\ref{eq44a})}{\longrightarrow}
\varphi_S\right|\times 
\stackrel{(\ref{eq23})}{(\chi_1,\chi_2)}
\stackrel{(\ref{eq14c},a)}{\longrightarrow}
(\phi_1,\phi_2)\right|\otimes (\psi_1,\psi_2)
\stackrel{(\ref{eq9})}{\longrightarrow}
\stackrel{(\ref{eq52p})}{{\mit \Psi}_S}
\nonumber\\
{\mit\Psi}_S&=&\psi_1
\left[(\vec{q}_1\cd \vec{\sigma}_{12})(\vec{q}_1\cd\vec{\sigma}_{3N})
-\frac{1}{3}\vec{q}\,_1^2 (\vec{\sigma}_{12}\cd\vec{\sigma}_{3N})\right.
+(\vec{q}_2\cd \vec{\sigma}_{12})(\vec{q}_2\cd\vec{\sigma}_{3N})
-\frac{1}{3}\vec{q}\,_2^2 (\vec{\sigma}_{12}\cd\vec{\sigma}_{3N})
\nonumber\\
&+&\quad\left.
(\vec{q}_3\cd \vec{\sigma}_{12})(\vec{q}_3\cd\vec{\sigma}_{3N})
-\frac{1}{3}\vec{q}\,_3^2 (\vec{\sigma}_{12}\cd\vec{\sigma}_{3N})\right]
(\vec{\tau}_{12}\cd\vec{\tau}_{3N}-3)
\nonumber\\ &+&\psi_2
\left[(\vec{q}_1\cd \vec{\sigma}_{12})(\vec{q}_1\cd\vec{\sigma}_{3N})
-\frac{1}{3}\vec{q}\,_1^2 (\vec{\sigma}_{12}\cd\vec{\sigma}_{3N})\right.
+(\vec{q}_2\cd \vec{\sigma}_{12})(\vec{q}_2\cd\vec{\sigma}_{3N})
-\frac{1}{3}\vec{q}\,_2^2 (\vec{\sigma}_{12}\cd\vec{\sigma}_{3N})
\nonumber\\
&+&\quad\left.
(\vec{q}_3\cd \vec{\sigma}_{12})(\vec{q}_3\cd\vec{\sigma}_{3N})
-\frac{1}{3}\vec{q}\,_3^2 (\vec{\sigma}_{12}\cd\vec{\sigma}_{3N})\right]
(\vec{\tau}_{12}\cd\vec{\tau}_{3N}+3)
\label{eq52p}\\
\hline
\nonumber\\
2\rightarrow 5:
&&
\left.\left.
\stackrel{(\ref{eq48},a)}{(f_1,f_2)}\times \stackrel{(\ref{eq44})}
{ S}
\stackrel{(\ref{eq44a})}{\longrightarrow}
(\varphi_1,\varphi_2)\right|\otimes 
\stackrel{(\ref{eq23})}{(\chi_1,\chi_2)}
\stackrel{(\ref{eq14c},c)}{\longrightarrow}
(\phi_1,\phi_2)\right|\otimes (\psi_1,\psi_2)
\stackrel{(\ref{eq9})}{\longrightarrow}
\stackrel{(\ref{eq53p})}{{\mit \Psi}_S}
\nonumber\\
{\mit\Psi}_S&=&\psi_1\left\{
\left[(\vec{q}_1\cd \vec{\sigma}_{12})(\vec{q}_1\cd\vec{\sigma}_{3N})
-\frac{1}{3}\vec{q}\,_1^2 (\vec{\sigma}_{12}\cd\vec{\sigma}_{3N})\right]
(\vec{\tau}_{12}\cd\vec{\tau}_{3N}-3)\right.
\nonumber\\
&-& 2\left[(\vec{q}_2\cd \vec{\sigma}_{12})(\vec{q}_2\cd\vec{\sigma}_{3N})
-\frac{1}{3}\vec{q}\,_2^2 (\vec{\sigma}_{12}\cd\vec{\sigma}_{3N})\right]
\vec{\tau}_{12}\cd\vec{\tau}_{3N} 
\nonumber\\
&+&
\left.
\left[(\vec{q}_3\cd \vec{\sigma}_{12})(\vec{q}_3\cd\vec{\sigma}_{3N})
-\frac{1}{3}\vec{q}\,_3^2 (\vec{\sigma}_{12}\cd\vec{\sigma}_{3N})\right]
(\vec{\tau}_{12}\cd\vec{\tau}_{3N}+3)\right\} 
\nonumber\\
&+&\psi_2\left\{
- 2\left[(\vec{q}_1\cd \vec{\sigma}_{12})(\vec{q}_1\cd\vec{\sigma}_{3N})
-\frac{1}{3}\vec{q}\,_1^2 (\vec{\sigma}_{12}\cd\vec{\sigma}_{3N})\right]
\vec{\tau}_{12}\cd\vec{\tau}_{3N} \right.
\nonumber\\
&+&
\left[(\vec{q}_2\cd \vec{\sigma}_{12})(\vec{q}_2\cd\vec{\sigma}_{3N})
-\frac{1}{3}\vec{q}\,_2^2 (\vec{\sigma}_{12}\cd\vec{\sigma}_{3N})\right]
(\vec{\tau}_{12}\cd\vec{\tau}_{3N}+3) 
\nonumber\\
&+&
\left.
\left[(\vec{q}_3\cd \vec{\sigma}_{12})(\vec{q}_3\cd\vec{\sigma}_{3N})
-\frac{1}{3}\vec{q}\,_3^2 (\vec{\sigma}_{12}\cd\vec{\sigma}_{3N})\right]
(\vec{\tau}_{12}\cd\vec{\tau}_{3N}-3)\right\}
\label{eq53p}\\
\hline
\nonumber\\
2\rightarrow 6:
&&
\left.\left.
\stackrel{(\ref{eq48},a)}{(f_1,f_2)}\times \stackrel{(\ref{eq44})}
{ S}
\stackrel{(\ref{eq44a})}{\longrightarrow}
(\varphi_1,\varphi_2)\right|\otimes 
\stackrel{(\ref{eq23})}{(\chi_1,\chi_2)}
\stackrel{(\ref{eq9cc})}{\longrightarrow}
\phi_S\right|\times \psi_S
\stackrel{(\ref{eq8})}{\longrightarrow}
\stackrel{(\ref{eq50p})}{{\mit \Psi}_S}
\nonumber\\
{\mit\Psi}_S&=&\psi_S\left\{
\left[(\vec{q}_1\cd \vec{\sigma}_{12})(\vec{q}_1\cd\vec{\sigma}_{3N})
-\frac{1}{3}\vec{q}\,_1^2 (\vec{\sigma}_{12}\cd\vec{\sigma}_{3N})\right]
(\vec{\tau}_{12}\cd\vec{\tau}_{3N}-3)\right.
\nonumber\\
&+&\quad\quad
\left[(\vec{q}_2\cd \vec{\sigma}_{12})(\vec{q}_2\cd\vec{\sigma}_{3N})
-\frac{1}{3}\vec{q}\,_2^2 (\vec{\sigma}_{12}\cd\vec{\sigma}_{3N})\right]
(\vec{\tau}_{12}\cd\vec{\tau}_{3N}+3) 
\nonumber\\
&-&\quad\,
\left.
2\left[(\vec{q}_3\cd \vec{\sigma}_{12})(\vec{q}_3\cd\vec{\sigma}_{3N})
-\frac{1}{3}\vec{q}\,_3^2 (\vec{\sigma}_{12}\cd\vec{\sigma}_{3N})\right]
\vec{\tau}_{12}\cd\vec{\tau}_{3N}\right\} 
\label{eq50p}\\
\hline
\nonumber\\
2\rightarrow 7:
&&
\left.\left.
\stackrel{(\ref{eq48},a)}{(f_1,f_2)}\times \stackrel{(\ref{eq44})}
{ S}
\stackrel{(\ref{eq44a})}{\longrightarrow}
(\varphi_1,\varphi_2)\right|\otimes 
\stackrel{(\ref{eq23})}{(\chi_1,\chi_2)}
\stackrel{(\ref{eq13c})}{\longrightarrow}
\phi_A\right|\times \psi_A
\stackrel{(\ref{eq21a})}{\longrightarrow}
\stackrel{(\ref{eq51p})}{{\mit \Psi}_S}
\nonumber\\
{\mit\Psi}_S&=&\psi_A\left\{
\left[(\vec{q}_1\cd \vec{\sigma}_{12})(\vec{q}_1\cd\vec{\sigma}_{3N})
-\frac{1}{3}\vec{q}\,_1^2 (\vec{\sigma}_{12}\cd\vec{\sigma}_{3N})\right]
(\vec{\tau}_{12}\cd\vec{\tau}_{3N}+1)\right.
\nonumber\\
&-&\quad
\left[(\vec{q}_2\cd \vec{\sigma}_{12})(\vec{q}_2\cd\vec{\sigma}_{3N})
-\frac{1}{3}\vec{q}\,_2^2 (\vec{\sigma}_{12}\cd\vec{\sigma}_{3N})\right]
(\vec{\tau}_{12}\cd\vec{\tau}_{3N}-1) 
\nonumber\\
&-&\left.
2\left[(\vec{q}_3\cd \vec{\sigma}_{12})(\vec{q}_3\cd\vec{\sigma}_{3N})
-\frac{1}{3}\vec{q}\,_3^2 (\vec{\sigma}_{12}\cd\vec{\sigma}_{3N})\right]
\right\} 
\label{eq51p}\\
\hline
\nonumber\\
3\rightarrow 5:
&&
\left.\left.
\stackrel{(\ref{eq48},b)}{(f'_1,f'_2)}\times \stackrel{(\ref{eq44})}
{ S}
\stackrel{(\ref{eq44a})}{\longrightarrow}
(\varphi_1,\varphi_2)\right|\otimes 
\stackrel{(\ref{eq23})}{(\chi_1,\chi_2)}
\stackrel{(\ref{eq14c},c)}{\longrightarrow}
(\phi_1,\phi_2)\right|\otimes (\psi_1,\psi_2)
\stackrel{(\ref{eq9})}{\longrightarrow}
\stackrel{(\ref{eq56p})}{{\mit \Psi}_S}
\nonumber\\
{\mit\Psi}_S&=&\psi_1\left\{
2\left[(\vec{q}_1\cd \vec{\sigma}_{12})(\vec{n}\cd\vec{\sigma}_{3N})
+(\vec{n}\cd \vec{\sigma}_{12})(\vec{q}_1\cd\vec{\sigma}_{3N})
-\frac{2}{3}(\vec{q}_1\cd\vec{n}) 
(\vec{\sigma}_{12}\cd\vec{\sigma}_{3N})\right]\right.
\nonumber\\
&+&\quad\left.
\left[(\vec{q}_2\cd \vec{\sigma}_{12})(\vec{n}\cd\vec{\sigma}_{3N})
+(\vec{n}\cd \vec{\sigma}_{12})(\vec{q}_2\cd\vec{\sigma}_{3N})
-\frac{2}{3}(\vec{q}_2\cd\vec{n}) 
(\vec{\sigma}_{12}\cd\vec{\sigma}_{3N})\right]
(\vec{\tau}_{12}\cd\vec{\tau}_{3N}+1)\right\}
\nonumber\\
&+&\psi_2\left\{ 
\left[(\vec{q}_1\cd \vec{\sigma}_{12})(\vec{n}\cd\vec{\sigma}_{3N})
+(\vec{n}\cd \vec{\sigma}_{12})(\vec{q}_1\cd\vec{\sigma}_{3N})
-\frac{2}{3}(\vec{q}_1\cd\vec{n})
(\vec{\sigma}_{12}\cd\vec{\sigma}_{3N})\right]
(\vec{\tau}_{12}\cd\vec{\tau}_{3N}-1)\right.
\nonumber\\
&-&2\left.
\left[(\vec{q}_2\cd \vec{\sigma}_{12})(\vec{n}\cd\vec{\sigma}_{3N})
+(\vec{n}\cd \vec{\sigma}_{12})(\vec{q}_2\cd\vec{\sigma}_{3N})
-\frac{2}{3}(\vec{q}_2\cd\vec{n})
(\vec{\sigma}_{12}\cd\vec{\sigma}_{3N})\right]\right\}
\label{eq56p}\\
\hline
\nonumber\\
3\rightarrow 6:
&&
\left.\left.
\stackrel{(\ref{eq48},b)}{(f'_1,f'_2)}\times \stackrel{(\ref{eq44})}
{ S}
\stackrel{(\ref{eq44a})}{\longrightarrow}
(\varphi_1,\varphi_2)\right|\otimes 
\stackrel{(\ref{eq23})}{(\chi_1,\chi_2)}
\stackrel{(\ref{eq9cc})}{\longrightarrow}
\phi_S\right|\times \psi_S
\stackrel{(\ref{eq8})}{\longrightarrow}
\stackrel{(\ref{eq54p})}{{\mit \Psi}_S}
\nonumber\\
{\mit\Psi}_S&=&\psi_S\left\{
\left[(\vec{q}_1\cd \vec{\sigma}_{12})(\vec{n}\cd\vec{\sigma}_{3N})
+(\vec{n}\cd \vec{\sigma}_{12})(\vec{q}_1\cd\vec{\sigma}_{3N})
-\frac{2}{3}(\vec{q}_1\cd\vec{n}) 
(\vec{\sigma}_{12}\cd\vec{\sigma}_{3N})\right]
(\vec{\tau}_{12}\cd\vec{\tau}_{3N}-1)\right.
\nonumber\\
&+&\quad\left.
\left[(\vec{q}_2\cd \vec{\sigma}_{12})(\vec{n}\cd\vec{\sigma}_{3N})
+(\vec{n}\cd \vec{\sigma}_{12})(\vec{q}_2\cd\vec{\sigma}_{3N})                                                       
-\quad\frac{2}{3}(\vec{q}_2\cd\vec{n}) 
(\vec{\sigma}_{12}\cd\vec{\sigma}_{3N})\right]
(\vec{\tau}_{12}\cd\vec{\tau}_{3N}+1)\right\}
\nonumber\\
&&
\label{eq54p}\\
\hline
\nonumber\\
3\rightarrow 7:
&&
\left.\left.
\stackrel{(\ref{eq48},b)}{(f'_1,f'_2)}\times \stackrel{(\ref{eq44})}
{ S}
\stackrel{(\ref{eq44a})}{\longrightarrow}
(\varphi_1,\varphi_2)\right|\otimes 
\stackrel{(\ref{eq23})}{(\chi_1,\chi_2)}
\stackrel{(\ref{eq13c})}{\longrightarrow}
\phi_A\right|\times \psi_A
\stackrel{(\ref{eq21a})}{\longrightarrow}
\stackrel{(\ref{eq55p})}{{\mit \Psi}_S}
\nonumber\\
{\mit\Psi}_S&=&\psi_A\left\{
\left[(\vec{q}_1\cd \vec{\sigma}_{12})(\vec{n}\cd\vec{\sigma}_{3N})
+(\vec{n}\cd \vec{\sigma}_{12})(\vec{q}_1\cd\vec{\sigma}_{3N})
-\frac{2}{3}(\vec{q}_1\cd\vec{n}) 
(\vec{\sigma}_{12}\cd\vec{\sigma}_{3N})\right]
(\vec{\tau}_{12}\cd\vec{\tau}_{3N}+3)\right.
\nonumber\\
&+&\quad\left.
\left[(\vec{q}_2\cd \vec{\sigma}_{12})(\vec{n}\cd\vec{\sigma}_{3N})
+(\vec{n}\cd \vec{\sigma}_{12})(\vec{q}_2\cd\vec{\sigma}_{3N})
-\frac{2}{3}(\vec{q}_2\cd\vec{n}) 
(\vec{\sigma}_{12}\cd\vec{\sigma}_{3N})\right]
(3-\vec{\tau}_{12}\cd\vec{\tau}_{3N})\right\}
\nonumber\\
&&
\label{eq55p}\\
\hline
\nonumber
\end{eqnarray}

\section{Wave functions generated by the antisymmetric tensor 
structures} \label{appwf2} 
As described in sect. \ref{asymm}, the antisymmetric tensors generate 
the functions given below.  Like in appendix \ref{appwf1}, the diagram 
shown below summarizes all the ways of generating the nucleon wave 
function ${\mit \Psi}_S$ from the functions of different symmetries.

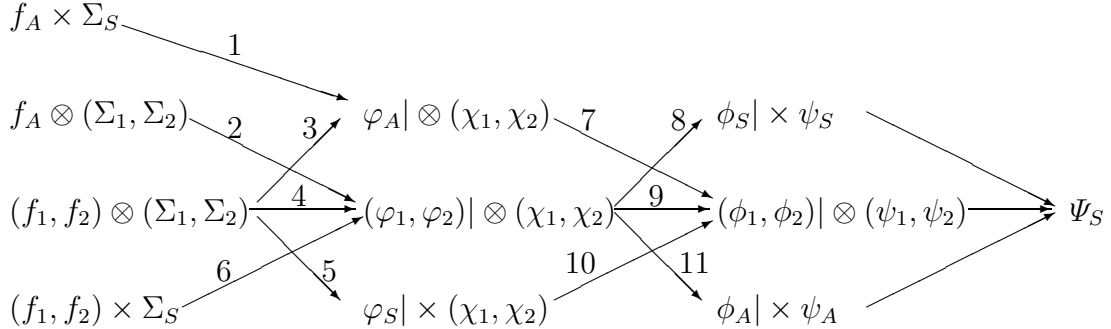
\begin{picture}(150,54)
\put(2,44){$f_A\times\Sigma_S$}
\put(2,31){$f_A\otimes (\Sigma_1,\Sigma_2)$}
\put(2,18){$(f_1,f_2)\otimes (\Sigma_1,\Sigma_2)$}
\put(2,5){$(f_1,f_2)\times \Sigma_S$}

\put(49,31){$\varphi_A|\otimes (\chi_1,\chi_2)$}
\put(49,18){$(\varphi_1,\varphi_2)|\otimes (\chi_1,\chi_2)$}
\put(49,5){$\varphi_S|\times (\chi_1,\chi_2)$}

\put(96,31){$\phi_S|\times\psi_S$}
\put(96,18){$(\phi_1,\phi_2)|\otimes (\psi_1,\psi_2)$}
\put(96,5){$\phi_A|\times\psi_A$}

\put(143,18){${\mit \Psi}_S$}

\put(17,44){\vector(3,-1){30}}
\put(26,31.5){\vector(2,-1){22}}
\put(35,20.5){\vector(1,1){11}}
\put(34,19.5){\vector(1,0){14}}
\put(35,18.5){\vector(1,-1){11}}
\put(25,6.5){\vector(2,1){24}}

\put(31,40){1}
\put(31,29){2}
\put(43,29){\llap{3}}
\put(39.5,20){4}
\put(43.5,10){5}
\put(29.5,10){6}

\put(74.5,31.5){\vector(2,-1){21}}
\put(82.5,20){\vector(1,1){11.5}}
\put(82.5,19.5){\vector(1,0){12.5}}
\put(82.5,19){\vector(1,-1){11.5}}
\put(74.5,7.5){\vector(2,1){21}}

\put(80,30){\llap{7}}
\put(92,30){\llap{8}}
\put(87,20){9}
\put(80,11){\llap{10}}
\put(91,11){11}

\put(116,32.5){\vector(2,-1){25}}
\put(129.5,19.5){\vector(1,0){11}}
\put(116,6.5){\vector(2,1){25}}
\end{picture}

For example, the path $1\rightarrow 7$ means that the antisymmetric 
function $f_A$ multiplied by the symmetric spin tensor $\Sigma_S$ gives 
the antisymmetric  $\varphi_A$. Then with the isospin doublet 
$(\chi_1,\chi_2)$, by eq. (\ref{eq14c},b), we construct the 
spin-isospin doublet $(\phi_1,\phi_2)$. Then with the momentum scalar 
functions $(\psi_1,\psi_2)$, by (\ref{eq9}), we get ${\mit \Psi}_S$ 
given by (\ref{eq69a}).  given by (\ref{eq53p}). Twelve paths 
connecting the initial states at the left of the diagram with the final 
one ${\mit \Psi}_S$ correspond to twelve functions given below.

As explained in appendix \ref{appwf1}, before any function we indicate 
in more detail, how the given structure is obtained. 
\begin{eqnarray} 
\hline
\nonumber\\
1\rightarrow 7:
&&
\left.\left.
\stackrel{(\ref{eq61a})}{f_A}\times \stackrel{(\ref{eq53da})}
{\Sigma_S}
\stackrel{(\ref{as1a},b)}{\longrightarrow}
\varphi_A\right|\otimes 
\stackrel{(\ref{eq23})}{(\chi_1,\chi_2)}
\stackrel{(\ref{eq14c},b)}{\longrightarrow}
(\phi_1,\phi_2)\right|\times (\psi_1,\psi_2)
\stackrel{(\ref{eq9})}{\longrightarrow}
\stackrel{(\ref{eq69a})}{{\mit \Psi}_S}
\nonumber\\
{\mit\Psi}_S&=&-\psi_1\{
(\vec{q}_1\cd\vec{\sigma}_{12})(\vec{q}_2\cd\vec{\sigma}_{3N})
-(\vec{q}_2\cd\vec{\sigma}_{12})(\vec{q}_1\cd\vec{\sigma}_{3N})
-2i\vec{\sigma}_{12}\cd[\vec{q}_1\times\vec{q}_2]\}
(1+\vec{\tau}_{12}\cd\vec{\tau}_{3N})
\nonumber\\
&-& \psi_2\{
(\vec{q}_1\cd\vec{\sigma}_{12})(\vec{q}_2\cd\vec{\sigma}_{3N})
-(\vec{q}_2\cd\vec{\sigma}_{12})(\vec{q}_1\cd\vec{\sigma}_{3N})
-2i\vec{\sigma}_{12}\cd[\vec{q}_1\times\vec{q}_2]\}
(1-\vec{\tau}_{12}\cd\vec{\tau}_{3N})
\label{eq69a}\\
\hline
\nonumber\\
2\rightarrow 8:
&&
\left.\left.
\stackrel{(\ref{eq61a})}{f_A}\otimes \stackrel{(\ref{eq53db})}
{(\Sigma_1,\Sigma_2)}
\stackrel{(\ref{as2},b)}{\longrightarrow}
(\varphi_1,\varphi_2)\right|\otimes 
\stackrel{(\ref{eq23})}{(\chi_1,\chi_2)}
\stackrel{(\ref{eq9cc})}{\longrightarrow}
\phi_S\right|\times \psi_S
\stackrel{(\ref{eq8})}{\longrightarrow}
\stackrel{(\ref{eq63a})}{{\mit \Psi}_S}
\nonumber\\
{\mit\Psi}_S&=&\psi_S 
\{(\vec{q}_1\cd\vec{\sigma}_{12})(\vec{q}_2\cd\vec{\sigma}_{3N})
-(\vec{q}_2\cd\vec{\sigma}_{12})(\vec{q}_1\cd\vec{\sigma}_{3N})
+i\vec{\sigma}_{12}\cd[\vec{q}_1\times\vec{q}_2]
\nonumber\\
&-&\quad i\vec{\sigma}_{3N}\cd[\vec{q}_1\times\vec{q}_2]
\vec{\tau}_{12}\cd\vec{\tau}_{3N}\}
\nonumber\\
&&
\label{eq63a}\\
\hline
\nonumber\\
2\rightarrow 9:
&&
\left.\left.
\stackrel{(\ref{eq61a})}{f_A}\otimes \stackrel{(\ref{eq53db})}
{(\Sigma_1,\Sigma_2)}
\stackrel{(\ref{as2},b)}{\longrightarrow}
(\varphi_1,\varphi_2)\right|\otimes 
\stackrel{(\ref{eq23})}{(\chi_1,\chi_2)}
\stackrel{(\ref{eq14c},c)}{\longrightarrow}
(\phi_1,\phi_2)\right|\times (\psi_1,\psi_2)
\stackrel{(\ref{eq9})}{\longrightarrow}
\stackrel{(\ref{eq72a})}{{\mit \Psi}_S}
\nonumber\\
{\mit\Psi}_S&=&\psi_1\{
[(\vec{q}_1\cd\vec{\sigma}_{12})(\vec{q}_2\cd\vec{\sigma}_{3N})
-(\vec{q}_2\cd\vec{\sigma}_{12})(\vec{q}_1\cd\vec{\sigma}_{3N})
+i\vec{\sigma}_{12}\cd[\vec{q}_1\times\vec{q}_2]]
(1-\vec{\tau}_{12}\cd\vec{\tau}_{3N})
\nonumber\\
&+&i\vec{\sigma}_{3N}\cd[\vec{q}_1\times\vec{q}_2]
(3+\vec{\tau}_{12}\cd\vec{\tau}_{3N})\}
\nonumber\\
&+&\psi_2\{
[(\vec{q}_1\cd\vec{\sigma}_{12})(\vec{q}_2\cd\vec{\sigma}_{3N})
-(\vec{q}_2\cd\vec{\sigma}_{12})(\vec{q}_1\cd\vec{\sigma}_{3N})
+i\vec{\sigma}_{12}\cd[\vec{q}_1\times\vec{q}_2]]
(1+\vec{\tau}_{12}\cd\vec{\tau}_{3N})
\nonumber\\
&-&i\vec{\sigma}_{3N}\cd[\vec{q}_1\times\vec{q}_2]
(3-\vec{\tau}_{12}\cd\vec{\tau}_{3N})\}
\label{eq72a}\\
\hline
\nonumber\\
2\rightarrow 11:
&&
\left.\left.
\stackrel{(\ref{eq61a})}{f_A}\otimes \stackrel{(\ref{eq53db})}
{(\Sigma_1,\Sigma_2)}
\stackrel{(\ref{as2},b)}{\longrightarrow}
(\varphi_1,\varphi_2)\right|\otimes 
\stackrel{(\ref{eq23})}{(\chi_1,\chi_2)}
\stackrel{(\ref{eq13c})}{\longrightarrow}
\phi_A\right|\times \psi_A
\stackrel{(\ref{eq21a})}{\longrightarrow}
\stackrel{(\ref{eq66a})}{{\mit \Psi}_S}
\nonumber\\
{\mit\Psi}_S&=&\psi_A
\{-\left[(\vec{q}_1\cd\vec{\sigma}_{12})(\vec{q}_2\cd\vec{\sigma}_{3N})
-(\vec{q}_2\cd\vec{\sigma}_{12})(\vec{q}_1\cd\vec{\sigma}_{3N})
+i\vec{\sigma}_{12}\cd[\vec{q}_1\times\vec{q}_2]\right]
\vec{\tau}_{12}\cd\vec{\tau}_{3N}
\nonumber\\
&-&\quad 3i\vec{\sigma}_{3N}\cd[\vec{q}_1\times\vec{q}_2]\}
\label{eq66a}\\
\hline
\nonumber\\
3\rightarrow 7:
&&
\left.\left.
\stackrel{(\ref{eq61b})}{(f_1,f_2)}\otimes \stackrel{(\ref{eq53db})}
{(\Sigma_1,\Sigma_2)}
\stackrel{(\ref{as1a},a)}{\longrightarrow}
\varphi_A\right|\otimes 
\stackrel{(\ref{eq23})}{(\chi_1,\chi_2)}
\stackrel{(\ref{eq14c},b)}{\longrightarrow}
(\phi_1,\phi_2)\right|\times (\psi_1,\psi_2)
\stackrel{(\ref{eq9})}{\longrightarrow}
\stackrel{(\ref{eq70a})}{{\mit \Psi}_S}
\nonumber\\
{\mit\Psi}_S&=&\psi_1\{
(\vec{q}_1\cd\vec{\sigma}_{12})(\vec{n}\cd\vec{\sigma}_{3N})
-(\vec{n}\cd\vec{\sigma}_{12})(\vec{q}_1\cd\vec{\sigma}_{3N})
-(\vec{q}_2\cd\vec{\sigma}_{12})(\vec{n}\cd\vec{\sigma}_{3N})
+(\vec{n}\cd\vec{\sigma}_{12})(\vec{q}_2\cd\vec{\sigma}_{3N})
\nonumber\\
&+&\quad i\vec{\sigma}_{12}\cd[\vec{q}_1\times\vec{n}]
+3i\vec{\sigma}_{3N}\cd[\vec{q}_1\times\vec{n}]
-i\vec{\sigma}_{12}\cd[\vec{q}_2\times\vec{n}]
+3i\vec{\sigma}_{3N}\cd[\vec{q}_2\times\vec{n}]\}
(1+\vec{\tau}_{12}\cd\vec{\tau}_{3N})
\nonumber\\
&+&\psi_2\{
(\vec{q}_1\cd\vec{\sigma}_{12})(\vec{n}\cd\vec{\sigma}_{3N})
-(\vec{n}\cd\vec{\sigma}_{12})(\vec{q}_1\cd\vec{\sigma}_{3N})
-(\vec{q}_2\cd\vec{\sigma}_{12})(\vec{n}\cd\vec{\sigma}_{3N})
+(\vec{n}\cd\vec{\sigma}_{12})(\vec{q}_2\cd\vec{\sigma}_{3N})
\nonumber\\
&+&\quad i\vec{\sigma}_{12}\cd[\vec{q}_1\times\vec{n}]
+3i\vec{\sigma}_{3N}\cd[\vec{q}_1\times\vec{n}]
-i\vec{\sigma}_{12}\cd[\vec{q}_2\times\vec{n}]
+3i\vec{\sigma}_{3N}\cd[\vec{q}_2\times\vec{n}]\}
(1-\vec{\tau}_{12}\cd\vec{\tau}_{3N})
\nonumber\\
&&
\label{eq70a}\\
\hline
\nonumber\\
4\rightarrow 8:
&&
\left.\left.
\stackrel{(\ref{eq61b})}{(f_1,f_2)}\otimes \stackrel{(\ref{eq53db})}
{(\Sigma_1,\Sigma_2)}
\stackrel{(\ref{as2},c)}{\longrightarrow}
(\varphi_1,\varphi_2)\right|\otimes 
\stackrel{(\ref{eq23})}{(\chi_1,\chi_2)}
\stackrel{(\ref{eq9cc})}{\longrightarrow}
\phi_S\right|\times \psi_S
\stackrel{(\ref{eq8})}{\longrightarrow}
\stackrel{(\ref{eq64a})}{{\mit \Psi}_S}
\nonumber\\
{\mit\Psi}_S&=&\psi_S\{
-[(\vec{q}_1\cd\vec{\sigma}_{12})(\vec{n}\cd\vec{\sigma}_{3N})
-(\vec{n}\cd\vec{\sigma}_{12})(\vec{q}_1\cd\vec{\sigma}_{3N})
+i\vec{\sigma}_{12}\cd[\vec{q}_1\times\vec{n}]]
(1+\vec{\tau}_{12}\cd\vec{\tau}_{3N})
\nonumber\\
&+&\quad [(\vec{q}_2\cd\vec{\sigma}_{12})(\vec{n}\cd\vec{\sigma}_{3N})
-(\vec{n}\cd\vec{\sigma}_{12})(\vec{q}_2\cd\vec{\sigma}_{3N})
+i\vec{\sigma}_{12}\cd[\vec{q}_2\times\vec{n}]]
(1-\vec{\tau}_{12}\cd\vec{\tau}_{3N})
\nonumber\\
&+&\quad i\vec{\sigma}_{3N}\cd[\vec{q}_1\times\vec{n}]
(3-\vec{\tau}_{12}\cd\vec{\tau}_{3N})
+i\vec{\sigma}_{3N}\cd[\vec{q}_2\times\vec{n}]
(3+\vec{\tau}_{12}\cd\vec{\tau}_{3N})\}
\label{eq64a}\\
\hline
\nonumber\\
4\rightarrow 9:
&&
\left.\left.
\stackrel{(\ref{eq61b})}{(f_1,f_2)}\otimes \stackrel{(\ref{eq53db})}
{(\Sigma_1,\Sigma_2)}
\stackrel{(\ref{as2},c)}{\longrightarrow}
(\varphi_1,\varphi_2)\right|\otimes 
\stackrel{(\ref{eq23})}{(\chi_1,\chi_2)}
\stackrel{(\ref{eq14c},c)}{\longrightarrow}
(\phi_1,\phi_2)\right|\times (\psi_1,\psi_2)
\stackrel{(\ref{eq9})}{\longrightarrow}
\stackrel{(\ref{eq73a})}{{\mit \Psi}_S}
\nonumber\\
{\mit\Psi}_S&=&\psi_1\{
[(\vec{q}_1\cd\vec{\sigma}_{12})(\vec{n}\cd\vec{\sigma}_{3N})
-(\vec{n}\cd\vec{\sigma}_{12})(\vec{q}_1\cd\vec{\sigma}_{3N})
+i\vec{\sigma}_{12}\cd[\vec{q}_1\times\vec{n}]]
(1+\vec{\tau}_{12}\cd\vec{\tau}_{3N})
\nonumber\\
&+&\quad
i\vec{\sigma}_{3N}\cd[\vec{q}_1\times\vec{n}]]
(3-\vec{\tau}_{12}\cd\vec{\tau}_{3N})
\nonumber\\
&+&\quad
2[(\vec{q}_2\cd\vec{\sigma}_{12})(\vec{n}\cd\vec{\sigma}_{3N})
-(\vec{n}\cd\vec{\sigma}_{12})(\vec{q}_2\cd\vec{\sigma}_{3N})
+i\vec{\sigma}_{12}\cd[\vec{q}_2\times\vec{n}]
-i\vec{\sigma}_{3N}\cd[\vec{q}_2\times\vec{n}]
\vec{\tau}_{12}\cd\vec{\tau}_{3N}]\}
\nonumber\\
&-&\quad\psi_2\{
[(\vec{q}_2\cd\vec{\sigma}_{12})(\vec{n}\cd\vec{\sigma}_{3N})
-(\vec{n}\cd\vec{\sigma}_{12})(\vec{q}_2\cd\vec{\sigma}_{3N})
+i\vec{\sigma}_{12}\cd[\vec{q}_2\times\vec{n}]]
(1-\vec{\tau}_{12}\cd\vec{\tau}_{3N})
\nonumber\\
&-&\quad
i\vec{\sigma}_{3N}\cd[\vec{q}_2\times\vec{n}]
(3+\vec{\tau}_{12}\cd\vec{\tau}_{3N})
\nonumber\\
&+&\quad 2[(\vec{q}_1\cd\vec{\sigma}_{12})(\vec{n}\cd\vec{\sigma}_{3N})
-(\vec{n}\cd\vec{\sigma}_{12})(\vec{q}_1\cd\vec{\sigma}_{3N})
+i\vec{\sigma}_{12}\cd[\vec{q}_1\times\vec{n}]
-i\vec{\sigma}_{3N}\cd[\vec{q}_1\times\vec{n}]
\vec{\tau}_{12}\cd\vec{\tau}_{3N}]\}
\nonumber\\
&&
\label{eq73a}\\
\hline
\nonumber\\
4\rightarrow 11:
&&
\left.\left.
\stackrel{(\ref{eq61b})}{(f_1,f_2)}\otimes \stackrel{(\ref{eq53db})}
{(\Sigma_1,\Sigma_2)}
\stackrel{(\ref{as2},c)}{\longrightarrow}
(\varphi_1,\varphi_2)\right|\otimes 
\stackrel{(\ref{eq23})}{(\chi_1,\chi_2)}
\stackrel{(\ref{eq13c})}{\longrightarrow}
\phi_A\right|\times \psi_A
\stackrel{(\ref{eq21a})}{\longrightarrow}
\stackrel{(\ref{eq67a})}{{\mit \Psi}_S}
\nonumber\\
{\mit\Psi}_S&=&\psi_A\{
-[(\vec{q}_1\cd\vec{\sigma}_{12})(\vec{n}\cd\vec{\sigma}_{3N})
-(\vec{n}\cd\vec{\sigma}_{12})(\vec{q}_1\cd\vec{\sigma}_{3N})
+i\vec{\sigma}_{12}\cd[\vec{q}_1\times\vec{n}]]
(3-\vec{\tau}_{12}\cd\vec{\tau}_{3N})
\nonumber\\
&-&\quad
[(\vec{q}_2\cd\vec{\sigma}_{12})(\vec{n}\cd\vec{\sigma}_{3N})
-(\vec{n}\cd\vec{\sigma}_{12})(\vec{q}_2\cd\vec{\sigma}_{3N})
+i\vec{\sigma}_{12}\cd[\vec{q}_2\times\vec{n}]]
(3+\vec{\tau}_{12}\cd\vec{\tau}_{3N})
\nonumber\\
&-&\quad 3i\vec{\sigma}_{3N}\cd[\vec{q}_1\times\vec{n}]
(1+\vec{\tau}_{12}\cd\vec{\tau}_{3N})
+3i\vec{\sigma}_{3N}\cd[\vec{q}_2\times\vec{n}]
(1-\vec{\tau}_{12}\cd\vec{\tau}_{3N})\}
\label{eq67a}\\
\hline
\nonumber\\
5\rightarrow 10:
&&
\left.\left.
\stackrel{(\ref{eq61b})}{(f_1,f_2)}\otimes \stackrel{(\ref{eq53db})}
{(\Sigma_1,\Sigma_2)}
\stackrel{(\ref{as1})}{\longrightarrow}
\varphi_S\right|\times 
\stackrel{(\ref{eq23})}{(\chi_1,\chi_2)}
\stackrel{(\ref{eq14c},a)}{\longrightarrow}
(\phi_1,\phi_2)\right|\times (\psi_1,\psi_2)
\stackrel{(\ref{eq9})}{\longrightarrow}
\stackrel{(\ref{eq68a})}{{\mit \Psi}_S}
\nonumber\\
{\mit\Psi}_S&=&-\psi_1\{
(\vec{q}_1\cd\vec{\sigma}_{12})(\vec{n}\cd\vec{\sigma}_{3N})
-(\vec{n}\cd\vec{\sigma}_{12})(\vec{q}_1\cd\vec{\sigma}_{3N})
+(\vec{q}_2\cd\vec{\sigma}_{12})(\vec{n}\cd\vec{\sigma}_{3N})
-(\vec{n}\cd\vec{\sigma}_{12})(\vec{q}_2\cd\vec{\sigma}_{3N})
\nonumber\\
&+&\quad i\vec{\sigma}_{12}\cd[\vec{q}_1\times\vec{n}]
-i\vec{\sigma}_{3N}\cd[\vec{q}_1\times\vec{n}]
+i\vec{\sigma}_{12}\cd[\vec{q}_2\times\vec{n}]
+i\vec{\sigma}_{3N}\cd[\vec{q}_2\times\vec{n}]\}
(3-\vec{\tau}_{12}\cd\vec{\tau}_{3N})
\nonumber\\
&+&\psi_2\{
(\vec{q}_1\cd\vec{\sigma}_{12})(\vec{n}\cd\vec{\sigma}_{3N})
-(\vec{n}\cd\vec{\sigma}_{12})(\vec{q}_1\cd\vec{\sigma}_{3N})
+(\vec{q}_2\cd\vec{\sigma}_{12})(\vec{n}\cd\vec{\sigma}_{3N})
-(\vec{n}\cd\vec{\sigma}_{12})(\vec{q}_2\cd\vec{\sigma}_{3N})
\nonumber\\
&+&\quad i\vec{\sigma}_{12}\cd[\vec{q}_1\times\vec{n}]
-i\vec{\sigma}_{3N}\cd[\vec{q}_1\times\vec{n}]
+i\vec{\sigma}_{12}\cd[\vec{q}_2\times\vec{n}]
+i\vec{\sigma}_{3N}\cd[\vec{q}_2\times\vec{n}]\}
(3+\vec{\tau}_{12}\cd\vec{\tau}_{3N})
\nonumber\\
&&
\label{eq68a}\\
\hline
\nonumber\\
6\rightarrow 8:
&&
\left.\left.
\stackrel{(\ref{eq61b})}{(f_1,f_2)}\times \stackrel{(\ref{eq53da})}
{\Sigma_S}
\stackrel{(\ref{as2},a)}{\longrightarrow}
(\varphi_1,\varphi_2)\right|\otimes 
\stackrel{(\ref{eq23})}{(\chi_1,\chi_2)}
\stackrel{(\ref{eq9cc})}{\longrightarrow}
\phi_S\right|\times \psi_S
\stackrel{(\ref{eq8})}{\longrightarrow}
\stackrel{(\ref{eq62a})}{{\mit \Psi}_S}
\nonumber\\
{\mit\Psi}_S&=&\psi_S\{-[(\vec{q}_1\cd\vec{\sigma}_{12})
(\vec{n}\cd\vec{\sigma}_{3N})
-(\vec{n}\cd\vec{\sigma}_{12})(\vec{q}_1\cd\vec{\sigma}_{3N})
-2i\vec{\sigma}_{12}\cd[\vec{q}_1\times\vec{n}]]
(1-\vec{\tau}_{12}\cd\vec{\tau}_{3N})
\nonumber\\
&+&\quad 
[(\vec{q}_2\cd\vec{\sigma}_{12})(\vec{n}\cd\vec{\sigma}_{3N}) 
-(\vec{n}\cd\vec{\sigma}_{12})(\vec{q}_2\cd\vec{\sigma}_{3N})
-2i\vec{\sigma}_{12}\cd[\vec{q}_2\times\vec{n}]]
(1+\vec{\tau}_{12}\cd\vec{\tau}_{3N})\}
\label{eq62a}\\
\hline
\nonumber\\
6\rightarrow 9:
&&
\left.\left.
\stackrel{(\ref{eq61b})}{(f_1,f_2)}\times \stackrel{(\ref{eq53da})}
{\Sigma_S}
\stackrel{(\ref{as2},a)}{\longrightarrow}
(\varphi_1,\varphi_2)\right|\otimes 
\stackrel{(\ref{eq23})}{(\chi_1,\chi_2)}
\stackrel{(\ref{eq14c},c)}{\longrightarrow}
(\phi_1,\phi_2)\right|\times (\psi_1,\psi_2)
\stackrel{(\ref{eq9})}{\longrightarrow}
\stackrel{(\ref{eq71a})}{{\mit \Psi}_S}
\nonumber\\
{\mit\Psi}_S&=&-\psi_1\{
2[(\vec{q}_1\cd\vec{\sigma}_{12})(\vec{n}\cd\vec{\sigma}_{3N})
-(\vec{n}\cd\vec{\sigma}_{12})(\vec{q}_1\cd\vec{\sigma}_{3N})
-2i\vec{\sigma}_{12}\cd[\vec{q}_1\times\vec{n}]]
\nonumber\\
&+&\quad
+[(\vec{q}_2\cd\vec{\sigma}_{12})(\vec{n}\cd\vec{\sigma}_{3N})
-(\vec{n}\cd\vec{\sigma}_{12})(\vec{q}_2\cd\vec{\sigma}_{3N})
-2i\vec{\sigma}_{12}\cd[\vec{q}_2\times\vec{n}]]
(1+\vec{\tau}_{12}\cd\vec{\tau}_{3N})\}
\nonumber\\
&+&\quad\psi_2\{
[(\vec{q}_1\cd\vec{\sigma}_{12})(\vec{n}\cd\vec{\sigma}_{3N})
-(\vec{n}\cd\vec{\sigma}_{12})(\vec{q}_1\cd\vec{\sigma}_{3N})
-2i\vec{\sigma}_{12}\cd[\vec{q}_1\times\vec{n}]]
(1-\vec{\tau}_{12}\cd\vec{\tau}_{3N})
\nonumber\\
&+&\quad
+2[(\vec{q}_2\cd\vec{\sigma}_{12})(\vec{n}\cd\vec{\sigma}_{3N})
-(\vec{n}\cd\vec{\sigma}_{12})(\vec{q}_2\cd\vec{\sigma}_{3N})
-2i\vec{\sigma}_{12}\cd[\vec{q}_2\times\vec{n}]]\}
\label{eq71a}\\
\hline
\nonumber\\
6\rightarrow 11:
&&
\left.\left.
\stackrel{(\ref{eq61b})}{(f_1,f_2)}\times \stackrel{(\ref{eq53da})}
{\Sigma_S}
\stackrel{(\ref{as2},a)}{\longrightarrow}
(\varphi_1,\varphi_2)\right|\otimes 
\stackrel{(\ref{eq23})}{(\chi_1,\chi_2)}
\stackrel{(\ref{eq13c})}{\longrightarrow}
\phi_A\right|\times \psi_A
\stackrel{(\ref{eq21a})}{\longrightarrow}
\stackrel{(\ref{eq65a})}{{\mit \Psi}_S}
\nonumber\\
{\mit\Psi}_S&=&\psi_A\{
[(\vec{q}_1\cd\vec{\sigma}_{12})(\vec{n}\cd\vec{\sigma}_{3N})
-(\vec{n}\cd\vec{\sigma}_{12})(\vec{q}_1\cd\vec{\sigma}_{3N})
-2i\vec{\sigma}_{12}\cd[\vec{q}_1\times\vec{n}]]
(3+\vec{\tau}_{12}\cd\vec{\tau}_{3N})
\nonumber\\
&+&\quad [(\vec{q}_2\cd\vec{\sigma}_{12})(\vec{n}\cd\vec{\sigma}_{3N})
-(\vec{n}\cd\vec{\sigma}_{12})(\vec{q}_2\cd\vec{\sigma}_{3N})
-2i\vec{\sigma}_{12}\cd[\vec{q}_2\times\vec{n}]]
(3-\vec{\tau}_{12}\cd\vec{\tau}_{3N})\}
\label{eq65a}\\
\hline
\nonumber
\end{eqnarray}


\end{document}